\documentclass[12pt,preprint]{aastex}

\newcommand{\kms}{\mbox{\,km\,s$^{-1}$}}

\newcommand{\Msun}{\,$M_{\odot}$}

\newcommand{\co}{\mbox{\rmfamily $^{12}$CO}}
\newcommand{\tco}{\mbox{\rmfamily $^{13}$CO}}
\newcommand{\ceont}{\mbox{\rmfamily C$^{18}$O}}
\newcommand{\ceo}{\mbox{\rmfamily C$^{18}$O}\,{(1--0)}}
\newcommand{\cso}{\mbox{\rmfamily C$^{17}$O}\,{(1--0)}}

\newcommand{\cod}{$c_{1D}$}
\newcommand{\vlsr}{$v_{LSR}$}

\newcommand{\sigmant}{$\sigma_{nt}$}
\newcommand{\degree}{$^{\circ}$}

\newcommand{\cmc}{${\rm cm}^{-3}$}
\newcommand{\hh}{H$_{2}$}

\newcommand{\tastar}{$T_{A}^{*}$}
\newcommand{\tmb}{$T_{mb}$}
\newcommand{\av}{$A_{V}$}
\newcommand{\tsys}{$T_{sys}$}

\newcommand{\sav}{$\sigma_{A_{V}}$}

\slugcomment{}
\shorttitle{Pipe Nebula: an improved CMF}
\shortauthors{Rathborne et al.}

\begin{document}

\title{Dense cores in the Pipe Nebula: An improved core mass function}
\author{J. M. Rathborne, C. J. Lada,  A. A. Muench}
\affil{Harvard-Smithsonian Center for Astrophysics, 60 Garden Street, Cambridge, MA 02138, USA: jrathborne@cfa.harvard.edu, clada@cfa.harvard.edu, gmuench@cfa.harvard.edu}
\and
\author{J. F. Alves}
\affil{Calar Alto Observatory, Centro Astron\'omico Hispano Alem\'an, c/Jes\'us Durb\'an Rem\'on 2-2, 04004, Almeria, Spain: jalves@caha.es}
\and
\author{J. Kainulainen}
\affil{TKK/Mets\"ahovi Radio Observatory, Mets\"ahovintie 114, FIN-02540 Kylm\"al\"a, Finland and Observatory, P.O. Box 14, FIN-00014 University of Helsinki, Finland: jouni.kainulainen@helsinki.fi}
\and
\author{M. Lombardi}
\affil{European Southern Observatory, Karl-Schwarzschild-Str. 2, 85748 Garching, Germany: mlombard@eso.org}
\begin{abstract}
In this paper we derive an improved core mass function (CMF) for the
Pipe Nebula from a detailed comparison between measurements of visual
extinction and molecular-line emission. We have compiled a refined
sample of 201 dense cores toward the Pipe Nebula using a
2-dimensional threshold identification algorithm informed by recent
simulations of dense core populations. Measurements of radial
velocities using complimentary \ceo\, observations enable us to cull
out from this sample those 43 extinction peaks that are either not associated with dense gas or are not
physically associated with the Pipe Nebula. Moreover, we use the
derived \ceont\, central velocities to differentiate between single
cores with internal structure and blends of two or more physically
distinct cores, superposed along the same line-of-sight. We then are
able to produce a more robust dense core sample for future
follow-up studies and a more reliable CMF than was possible
previously.  We confirm earlier indications that the CMF for the Pipe
Nebula departs from a single power-law like form with a break or knee
at M $\sim$ 2.7 $\pm$ 1.3\,\Msun. Moreover, we also confirm
that the CMF exhibits a similar shape to the stellar IMF, but is
scaled to higher masses by a factor of $\sim$ 4.5. We
interpret this difference in scaling to be a measure of the star
formation efficiency (22 $\pm$ 8\%). This supports
earlier suggestions that the stellar IMF may originate more or less
directly from the CMF.
\end{abstract}

\section{Introduction}

Recent studies of dense cores suggest that the fundamental mass
distribution of stars may be set during the early stages of core
formation. This is primarily because of the similarities between the
slopes of the stellar initial mass function (IMF) and those of the
observed core mass functions (CMFs) for core masses $>$ 1\,\Msun\,
(e.g.\,
\citealp{Motte98,Testi98,Johnstone00,Johnstone01,Motte01,Johnstone06a,Johnstone06b,Stanke06,Reid06a,Reid06b,Young06,Alves07,Nutter07,Enoch08,Simpson08}). If
the stellar IMF is in fact predetermined by the form of the CMF, then
the origin of the stellar IMF may be directly linked to the origin of
dense cores. Thus, understanding the connection between the CMF and
the stellar IMF is essential for theories of star formation.

Power-laws are by nature scale-free and the apparent similarity of the
CMF and IMF slopes does not necessarily imply a one-to-one
correspondence between the CMF and the IMF (e.g.,
\citealp{Swift08}). However, because the shape of the stellar IMF
flattens and departs from a single power-law at low masses (e.g.,
\citealp{Kroupa01}), any similar flattening measured for a CMF would
strengthen the idea of a very direct connection between the two
distributions. Moreover, the characteristic mass, that is, the mass at
which the distribution departs from a single power-law form, provides
a definite physical scaling for a mass function and differences in the
characteristic masses of the CMF and IMF would be a measure of
the efficiency of star formation (e.g., \citealp{Alves07}).
  
The reliability of any detailed statements made about the shape of the
CMF, or its connection to the stellar IMF, depends sensitively on the
uncertainties involved in the generation of the CMF. Small number
statistics, core crowding and the (in)accuracy of the core masses can
very easily introduce significant uncertainties in the derived shape
of the CMF (e.g., \citealp{Jouni}).  Even the operational definition
of a core can be a troubling source of uncertainty in the
determination of a CMF (e.g., \citealp{Swift08,Smith08}) .

To minimize such uncertainties and to obtain a meaningful measurement
of a CMF we first require a large sample of dense cores with accurate
measurements of their properties. Ideally we desire a sample of
starless cores in a single molecular cloud. Fortunately, the Pipe
Nebula, at a distance of 130 pc, is well suited for this purpose. The
cloud exhibits little evidence for active star formation and a
detailed dust extinction map of the complex exists containing numerous
core-like structures whose masses can be precisely measured
\citep{Lombardi06,Alves07}. Indeed, \cite{Alves07} identified 159
cores with masses between 0.2--20.0 \Msun\ in the Pipe
Nebula. Subsequent molecular-line observations demonstrated that many
of these were in fact dense cores (n(H$_2$) $>$ 10$^4$ cm$^{-3}$;
\citealp{Rathborne-pipe}) .  The CMF derived for this core
population was not a single power-law function. Instead, the CMF
exhibited a clear break at a mass of $\sim$ 2\,\Msun, suggesting that an
overall efficiency of about 30\% would characterize the star formation process
if these cores evolve to make stars \citep{Alves07}.

\cite{Jouni} performed simulations of the Pipe core population
and investigated the sensitivity of the derived CMF to the details of
the core extraction algorithm that was employed by \cite{Alves07}. The
simulations confirmed the presence of the break in the derived CMF and
revealed that the mass range where the break occurs is not affected by
either incompleteness or biases introduced by the extraction
algorithm. However, they found that the fidelity of the overall shape
of the CMF did depend on the choice of input parameters of the core
extraction algorithm and they determined optimum values for the search
parameters. These parameters differed somewhat from those used by
\cite{Alves07}.  Moreover, the simulations also indicated that
extractions sometimes produce spurious cores. Finally there are also
uncertainties inherent in the use of 2-dimensional data such as
extinction maps, to accurately define a core population. In
particular, such data cannot distinguish between any foreground or
background extinction features or cores. In addition it is difficult
to distinguish physically related substructure within a single object
from a blend of physically distinct cores along a similar
line-of-sight. These considerations suggested that a re-analysis of
the Pipe CMF was warranted, especially given the ramifications of the
results for core and star formation.

To derive a more reliable CMF for the Pipe Nebula and investigate its
relation to the stellar IMF, we have undertaken a study of cores
within the Pipe Nebula using a combination of visual extinction and
new \ceo\, molecular line data.  Because the core properties are
derived directly from how cores are defined and extracted, one needs
to pay careful attention to how this is done.  Here we use the visual
extinction image and a 2-dimensional clump finding algorithm, guided
by the \cite{Jouni} simulations, to identify and extract discrete
extinction features as candidate cores. The \ceo\, emission is then
used to determine if the candidate cores are (1) associated with dense gas, (2) associated with the Pipe Nebula, and (3)
separate or physically related structures. Because 2-dimensional
finding algorithms can pick up spurious or unrelated cores, we use
\ceo\, emission to filter the sample to include only real dense cores
that are associated with the molecular cloud.  This is particularly
important because the inclusion of spurious or unrelated cores, as
well as the artificial merging or separation of extinction features,
will significantly affect the shape of the CMF and any conclusions
derived from it.  With a reliable CMF we can then make detailed
comparisons between it and the stellar IMF to gain a better
understanding of the connection between dense cores and the star
formation process.

\section{Identification of discrete extinction peaks}
\label{clfind}

The identification of discrete extinction peaks within the Pipe
Nebula was performed using the visual extinction (\av) image of
\cite{Lombardi06}. This image was generated using the NICER method
which utilized the JHK photometry of over 4.5 million stars within the
2MASS catalog. In addition to the many compact extinction features,
this image also contains significant and variable extinction from the
lower-column density material associated with the molecular cloud. To
reliably extract discrete extinction peaks a background subtraction
was performed using a wavelet decomposition which filters out
structures larger than the specified size scale (0.30pc; see
\citealp{Alves07} for details).  This procedure produces a smoothly
varying background extinction image and a `cores-only' image. The
extinction features were extracted from the cores-only,
background-subtracted extinction image. All core parameters
(coordinates, peak \av, radius, and mass) were measured directly from
this cores-only map.

To identify the discrete extinction peaks, we use the 2-dimensional
version of the automated algorithm clumpfind (clfind2d;
\citealp{Williams94}). Clfind2d searches through an image using
iso-brightness surfaces to identify contiguous emission features
without assuming an a priori shape. The iso-brightness surfaces are
identified through a series of contour levels. In contrast to the
3-dimensional version of the algorithm, clfind2d allows users to input
arbitrary levels to determine the contouring. The algorithm starts its
search for emission features at the pixel with the peak brightness in
the image and steps down from this using the user-defined contour
levels.  During each iteration, the algorithm finds all contiguous
pixels between each particular contour level and the next level
down. If the contiguous pixels are isolated from any previously
identified emission features, then they are assigned to a new
clump. If they are connected they are assigned to a pre-existing
clump.  For cases of isolated emission features, this procedure is
straightforward. However, in the case of blended emission features, a
`friends-of-friends' algorithm (see \citealp{Williams94} for more
details) is used to separate the emission. The algorithm iterates
until the lowest contour level is reached. At all levels clfind2d
requires that each new clump is greater than a specified number of
pixels. To be conservative, we set this minimum number of pixels to be
20. Assuming this area is a circle, then this size corresponds to 2.5
times the angular resolution of the visual extinction image.

The original list of cores toward the Pipe Nebula was compiled by
\cite{Alves07} using the same wavelet decomposition and clfind2d 
algorithm described above. They identified 159 cores using extinction
contour levels of 1.2, 4.0, and 6.0 mags. Because the extinction
contouring was truncated at 6 magnitudes, many of the higher
extinction regions were considered to be single extinction features
when in fact some consist of multiple well separated extinction peaks.

To improve on the previous list of cores and to identify them in a
homogeneous way at all emission levels, we choose to define the
contour levels input into clfind2d using discreet values of the small
scale variation in the background (\sav). This was estimated by
calculating the standard deviation in the extinction values in $\sim$1
pc$^{2}$ regions across the background map. Because the extinction
which results from the larger-scale molecular cloud varies across the
Pipe Nebula, we use the value of \sav\, to be 0.4\,mags. This value is
higher than the individual noise in each pixel (0.2\,mags;
\citealp{Lombardi06}).

Our choice of clumpfind input parameters is based on the recent
simulations of \cite{Jouni}. To best match the real data, these
simulations placed elliptical Gaussian cores of various masses on the
background image of the Pipe Nebula using the observed median core
separation and column density distribution of \cite{Alves07}. Similar
to the real data, a background subtraction was performed using a
wavelet decomposition. The extraction of cores was then performed on
the background-subtracted image using a range of clumpfind contour
levels. These simulations were investigating the effect different
clumpfind contour levels have on the derived core properties and the
recoverability and completeness of the resulting CMF. While
\cite{Jouni} find that the exact choice of contour levels has little
effect on the overall mass completeness limit, the derived CMF is
slightly more complete when using clumpfind contours that include the
extinction values down to the image noise. More importantly, however,
these simulations show that the degree of crowding within a molecular
cloud can significantly effect both the measured core parameters and
the derived CMF.

The simulations reveal that the Pipe CMF is 90\% complete to a mass of
0.5\,\Msun\, when using contour levels starting at 0.4 mag (1\sav) and
increasing in steps of 1.2 mag (3\sav). In a typical simulation where
$\sim$230 cores were input, these clumpfind parameters extracted
$\sim$190 cores. While the number of extracted cores is close to the
number input, a visual inspection of the positions of the input and
extracted cores reveal significant differences.  We found that
$\sim$60\% of the extracted cores match well with those input, while
$\sim$33\% of the input cores were not extracted. More concerning,
however, is the large number of spurious cores that were extracted:
$\sim$ 40\% of those extracted did not match an input core. Because
these are typically low mass cores, it is likely that the majority of
these spurious cores may have been introduced by the wavelet
decomposition and are not real dense cores. Moreover, the simulations
also reveal that the process of wavelet decomposition filtered out
some of the input cores.  While these effects are most obvious for
cores that are below the mass completeness limit, the use of such a
low contour level artificially merges some of these noise fluctuations
with `real' cores, thereby increasing their mass which will then alter
the shape of the resulting CMF.

To reduce the number of these spurious cores extracted from the real
data, we chose to start the contouring at 1.2 mag (3\sav) rather than
0.4 mag (1\sav). Specifically, we use contour levels starting at 1.2
mag (3\sav) and increasing in steps of 1.2 mag (3\sav), i.e. 1.2, 2.4,
3.6, 4.8, 6.0, 7.2, 8.4, 9.6, 10.8, 12.0, 13.2, and 14.4 mags.

With these contour levels \cite{Jouni} find that from twenty
realizations of the simulation, the mean number of cores extracted is
$\sim$174 (when $\sim$240 are input). From the extracted cores in a
single simulation, $\sim$ 150 of them are coincident with an input
core, while $\sim$ 26 of them have no corresponding input core and
are, thus, spurious detections. The simulations reveal that the mass
completeness limit is $\sim$0.65\,\Msun. Most (97\%) of the input
cores that were not extracted were below this mass completeness
limit. Moreover, the most massive core missed was typically not
greater than $\sim$ 1.5\,\Msun. In addition, \cite{Jouni} find that
only about 15\% of the input cores are blended with their nearest
neighbor.

Using contour levels starting at 1.2 mag (3\sav) and increasing in
steps of 1.2 mag (3\sav) on the observed visual extinction cores-only
image, we extract 201 discrete extinction peaks from the real
data\footnote{These parameters appear to characterize the underlying
data well, the exception being toward the densest region in the Pipe
Nebula, Barnard 59. Using these parameters clfind2d identifies 6
extinction peaks toward Barnard 59.  Because of the high extinction
and incomplete number of stars toward this region in the 2MASS image,
the extinction image has artificial structure which clfind2d breaks up
in to multiple peaks. Higher-angular resolution images show, however,
that this core though mostly smooth may in fact contain substructure
consisting of 2--3 extinction peaks at its center.  For this work,
however, we incorporate all the extinction within this region in to a
single core.}. We use this list for all further analysis.

\section{Molecular line observations}

Observations of \ceo\, molecular line emission toward all 201
extinction peaks identified toward the Pipe Nebula were obtained using
the 12\,m Arizona Radio Observatory (ARO) and the 22\,m Mopra
telescopes. The ARO \ceo\, observations were conducted over two
periods in 2005--2006. In total, 102 of the 201 extinction peaks were
observed with the ARO.  Ninety-four of these positions corresponded to
cores within the original list of \cite{Alves07}. The details of these
observations are presented in \cite{Gus-pipe}.

Spectra were obtained with the Mopra telescope toward all remaining
identified extinction peaks for which we had no previous \ceo\,
data. In total, an additional 99 positions were observed over the
periods 2007 July 3--8 and 2008 July 20--25.  In all cases the spectra
were obtained toward the peak extinction identified using clfind2d and
the 2MASS visual extinction image ($\Theta_{HPBW}$ $\sim$ 1\arcmin).

For the Mopra observations the 8\,GHz spectrometer MOPS was used to
simultaneously observe the transitions \co, \tco, \ceont, and \cso.
The spectrometer was used in `zoom' mode such that one spectral window
covered each of the lines of interest. Each window was 137.5 MHz wide
and contained 4096 channels in both orthogonal polarizations.  This
produced a velocity resolution of 0.09\,\kms\, for the \ceo\,
transition. At these frequencies the Mopra beam is $\sim$ 33\arcsec\,
\citep{Ladd05}.

All spectra were obtained as four 5 minute integrations in the
position switched mode. A common `off' position was used in all cases
and was selected to lie within a region of the Pipe Nebula which is
devoid of cores and molecular gas ($\alpha$=17:22:51.75,
$\delta$=$-$25:24:12.39, J2000).  The telescope pointing was checked
approximately every hour using a suitably bright, nearby maser.

The spectra were reduced using the ATNF Spectral Analysis Package
(ASAP) and were initially baselined subtracted before averaged using a
system temperature (\tsys) weighting. The typical \tsys\, for these
observations was $\sim$ 320 K. Gaussian profiles were fit to each
spectrum to determine its peak temperature (\tastar), central velocity
(\vlsr), line-width ($\Delta$V), and integrated intensity (I).  To be
considered a significant detection, we require that the integrated
intensity be greater than 3 times the \tastar\, rms noise. All raw
data are in the \tastar\, scale. The final spectra have a typical
\tastar\, rms noise of $\sim$0.02\, K channel$^{-1}$.  To convert to the main
beam brightness, \tmb, one needs to use the main beam efficiency of
0.43 as listed in \cite{Ladd05}. Although the spectra have a velocity
resolution of $\sim$0.09\,\kms, the typical error on the measurement
of \vlsr\, from the Gaussian fits were $<$0.01\,\kms.

\section{Results}
\label{results}

Having identified a large number of extinction peaks within the
background-subtracted extinction image, we now use the \ceo\, emission
to determine if they are associated with dense ($\sim$10$^{3}$\cmc) gas and which are
associated with the Pipe molecular cloud. Excluding any spurious
extinction peaks that may have been artificially included via the
wavelet decomposition and any that are unrelated to the region is
crucial in generating a reliable CMF. In this section we also use the
\ceo\, emission to help determine whether nearby extinction features
are physically separate or part of the same physical core. With these
distinctions we can then derive reliable core ensemble properties
such as radius, mass, density, and non-thermal line-width.

\subsection{Dense gas associated with the extinction peaks}

Although we also have \co\, and \tco\, toward a large number of the
extinction peaks within the Pipe Nebula, we choose to focus on the
\ceont\, emission.  The critical density for excitation of \ceo\, is
6$\times$10$^{3}$\,\cmc\, and because the isotope is rarer and the
emission is optically thin, it is a far superior tracer of dense gas
within a core than either \co\, or \tco.  Combining the ARO and Mopra
\ceo\, observations, we find that 93\% of the extinction peaks have
significant \ceont\, molecular line emission.  Thus, we find that 188
of the extinction peaks are associated with dense gas.

The extinction peaks with no detected \ceo\, emission have low peak
extinctions (\av\,$<$ 3.6 mags) and low masses (M $<$ 1.3\,\Msun).
Moreover, the majority of these are isolated features and are located
toward the edges of the extinction image.  It is likely that these
extinction peaks are not real dense cores and could simply be noise
fluctuations artifically added in by the wavelet decomposition.

\subsection{Association with the Pipe Molecular Cloud}

\begin{figure}
\center
\includegraphics[angle=90,width=0.5\textwidth]{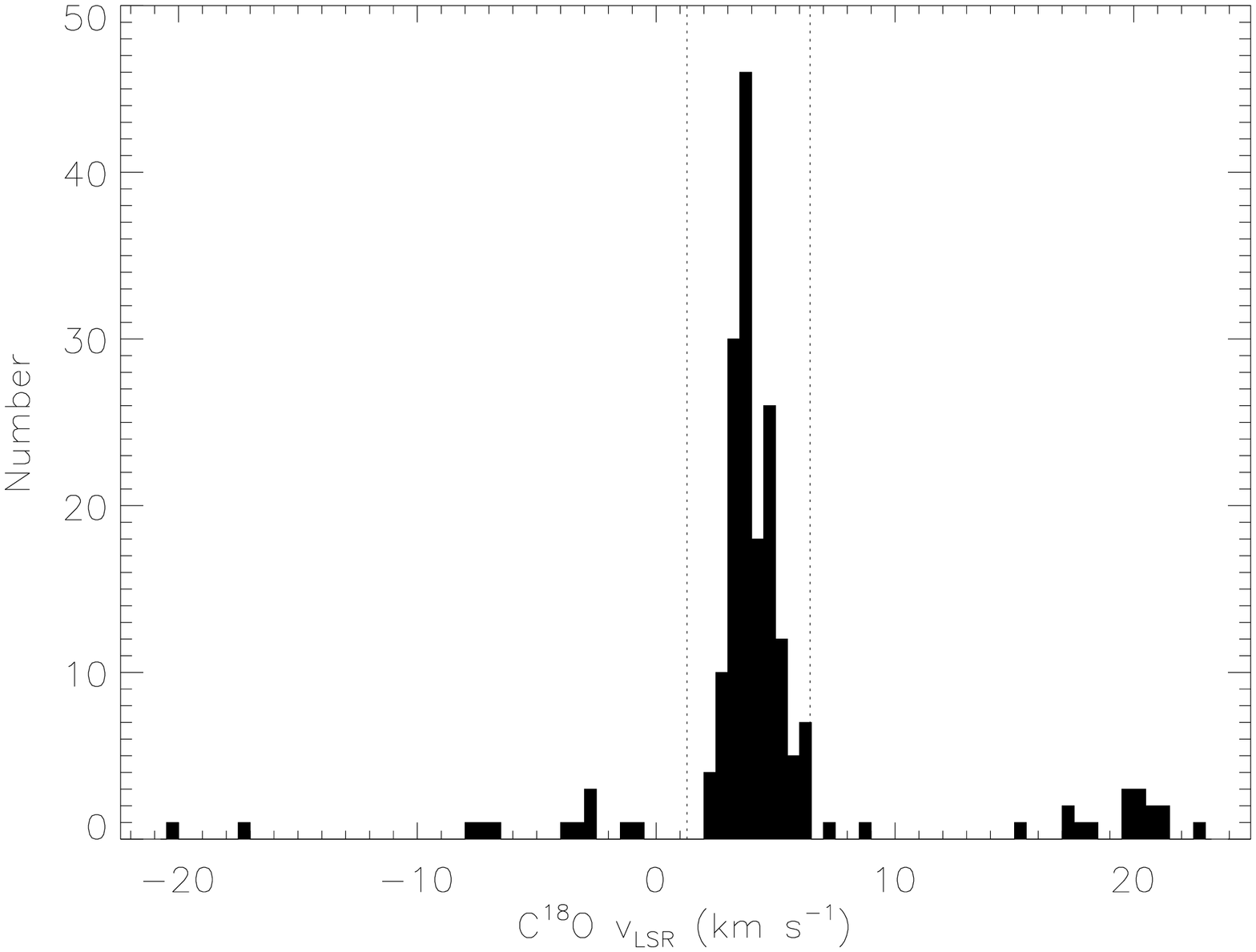}
\caption{\label{velocities} Measured \ceo\, central velocities (\vlsr) toward 
    the extinction peaks. The dotted vertical lines mark the range of
    velocities we define to determine which extinction peaks are
    associated with the Pipe Nebula. These values of 1.3\,\kms\, and
    6.4\,\kms, were determined from a Gaussian fit to the \vlsr\,
    distribution. Of the 188 extinction peaks that are associated with
    \ceo\, emission, we find that 158 of them have velocities in the
    range of 1.3\,\kms $<$ \vlsr $<$ 6.4\,\kms. We consider these
    extinction peaks to be associated with the Pipe molecular
    cloud. All others are foreground and/or background molecular cores
    and are not included within the final list.}
\end{figure}

Previous large-scale \tco\, observations show that the main body of
the Pipe Nebula is associated with molecular line emission in the
velocity range of 2\,\kms $<$ \vlsr $<$ 8\,\kms\,
\citep{Onishi99}. Figure~\ref{velocities} shows the measured \ceont\,
central velocities (\vlsr) toward all the extinction peaks with
significant \ceont\, emission. We find that the emission from the 188
extinction peaks falls within a narrow range.  A Gaussian fit to the
distribution reveals that it is centered at a \vlsr\, of $\sim$
3.9\,\kms\, with a FWHM of $\sim$ 2\,\kms.

To determine which extinction peaks are associated with the Pipe
Nebula, we use the range of 1.3\,\kms\, $<$ \vlsr\, $<$ 6.4\,\kms\,
(center \vlsr\, $\pm$ 3 times the FWHM). Of the 188 extinction peaks
that are associated with \ceont\, emission, we find that 158 of them
have velocities in this range. Thus, we will consider only these
extinction peaks as associated with the Pipe molecular cloud.

We find that 30 of the extinction peaks have associated molecular gas
that differs significantly from this velocity range, however. This
gas, at velocities of $\sim$$-$5 to 0\,\kms\, and $\sim$15 to
20\,\kms, is likely associated with foreground and/or background
molecular clouds. Indeed, there are 16 extinction peaks with a \vlsr\,
of $\sim$ 20\,\kms, that are localized in Galactic longitude and
latitude ($\ell \sim$3\,\degree, $b \sim$ 4.2\,\degree). These
extinction peaks are most likely related to a common background
molecular cloud.

For all further analysis, we consider only the 158 extinction peaks
that have dense gas at velocities associated with the Pipe
molecular cloud.

\subsection{Multiple, nearby extinction peaks: superimposed cores or cores with substructure?}

If cores are centrally condensed objects that will give rise to a
single star, then their extinction profiles should reflect their
density gradients and be approximately symmetric, centrally peaked,
and well defined above the background.  While many isolated extinction
features within the Pipe Nebula have these simple characteristics and
can be identified as discrete cores, there are several extinction
features that show more complex structure.

Because of the potential influence of crowding, many of these complex
extinction structures may actually comprise multiple cores. With the
2-dimensional visual extinction image alone, it is impossible to
determine if these complex structures comprise individual cores that
are superimposed along the line of sight or if they are single cores
with substructure that may not necessarily give rise to separate
star-forming events. Because we are using 2-dimensional extinction
data to determine the core properties, we need to distinguish between
these scenarios. An incorrect extraction of the core parameters will
directly influence their derived masses and, hence, the shape of the
resulting CMF. For example, the erroneous merging of many extinction
features into a single structure will artificially increase the number
of high mass cores and decrease the number of low mass cores. In
contrast, if the substructure within a larger core is artifically
separated into many individual objects, then there will be a
deficiency of high mass cores and an abundance of lower mass cores. In
either case, the derived CMF will be significantly affected.

Because we are interested in deriving reliable masses and, hence, the
CMF for cores within the Pipe Nebula, we need to distinguish between
extinction features that are the superposition of physically unrelated
cores from those that may have internal substructure. As a first
attempt at distinguishing between these scenarios, we use the \vlsr\,
derived from the \ceo\, emission. Because most ($\sim$80\%) of the
extinction peaks are well-defined isolated features, it is only
necessary to consider those larger extinction features that appear to
contain substructure. Thus, for this analysis, we consider only
multiple extinction peaks that happen to lie within a larger structure.

We define the larger extinction structures as contiguous regions of
visual extinctions greater than \,3$\sigma$ ($>$\,1.2\,mags) in the
background-subtracted image.  Given the clumpfind contouring levels
described in \S~\ref{clfind}, we find that 28 large extinction
structures contain multiple extinction peaks.  Because we have
measured \ceo\, emission toward all the extinction peaks, we can
calculate the difference in the \vlsr\, of each peak with respect to
all other peaks within the larger structure \footnote{The molecular
line emission associated with the isolated extinction peaks all
correspond to a single Gaussian \ceo\, line profile. Some positions
within the Pipe molecular ring \citep{Gus-pipe}, however, have
multiple \ceo\, line profiles. In these cases we consider all emission
lines when determining the relative \vlsr\, between neighboring
extinction peaks.}. If the velocity difference ($\delta$V) between two
extinction peaks is greater than the 1-dimensional projected sound
speed in a 10\,K gas (\cod\, = 0.12\,\kms), then we assume that the
\ceo\, emission is arising from physically different cores that happen
to be adjacent because they are superimposed along the line of
sight. On the other hand, if the velocity difference is less than
\cod, then the extinction peaks may in fact be physically
associated. To be considered part of the same physical core, we also
require that the separation between the extinction peaks with
velocities less than \cod\, is smaller than its Jeans length,
$R_{J}$. The Jeans length was calculated via the expression

\[ R_{J}  = \sqrt{\frac{15 k T_G}{8 \pi G m_{p} \rho}} \]

\noindent where $k$ is the Boltzmann constant, $T_{G}$ is the 
gas temperature, $G$ is the gravitational constant, $m_{p}$ is the
mass of a proton, and $\rho$ is the mass density. In all cases we
assume T$_{G}$=10\,K.

\begin{figure}
\center
\includegraphics[height=0.25\textheight,clip=true]{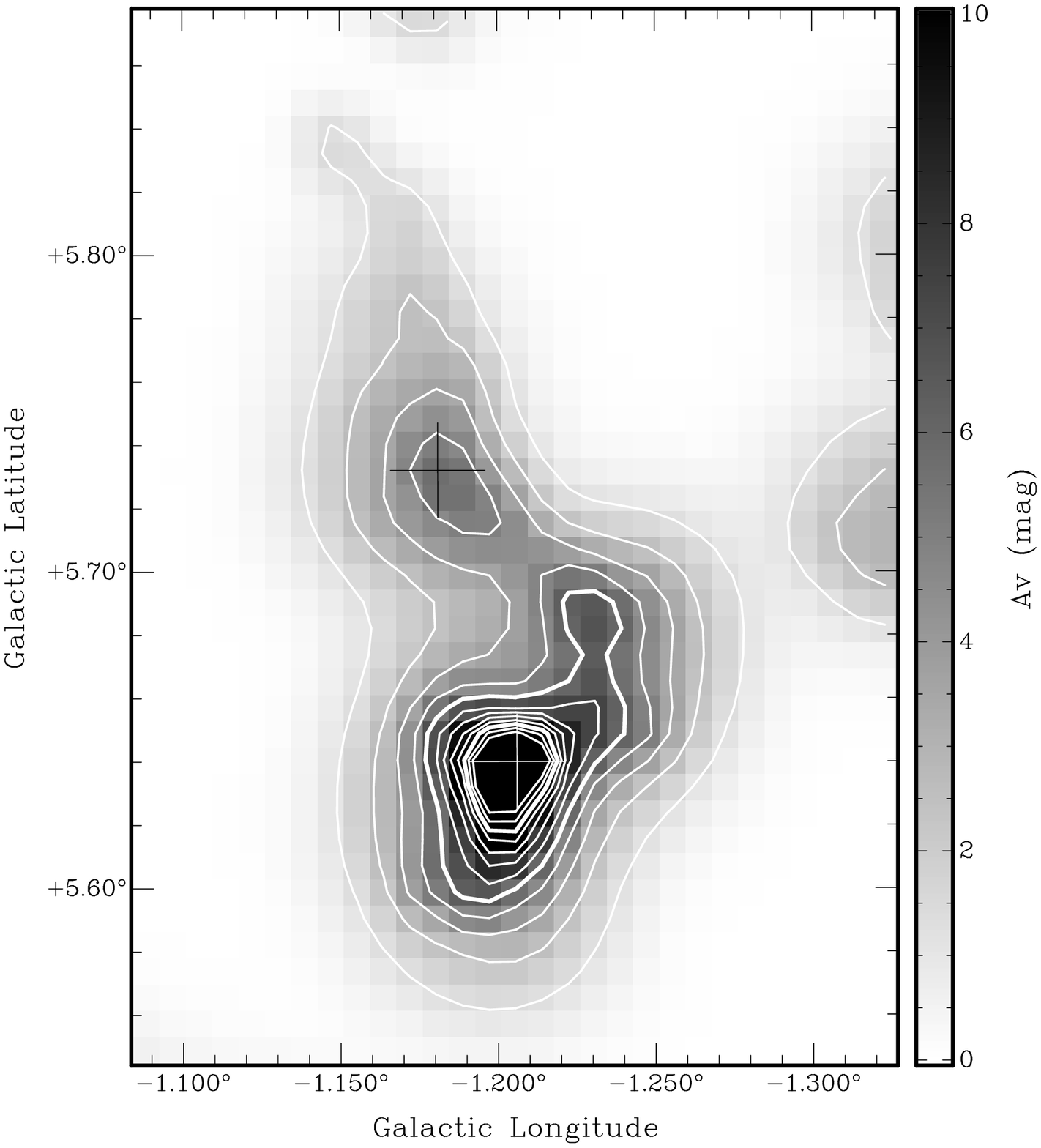} \includegraphics[height=0.25\textheight,clip=true]{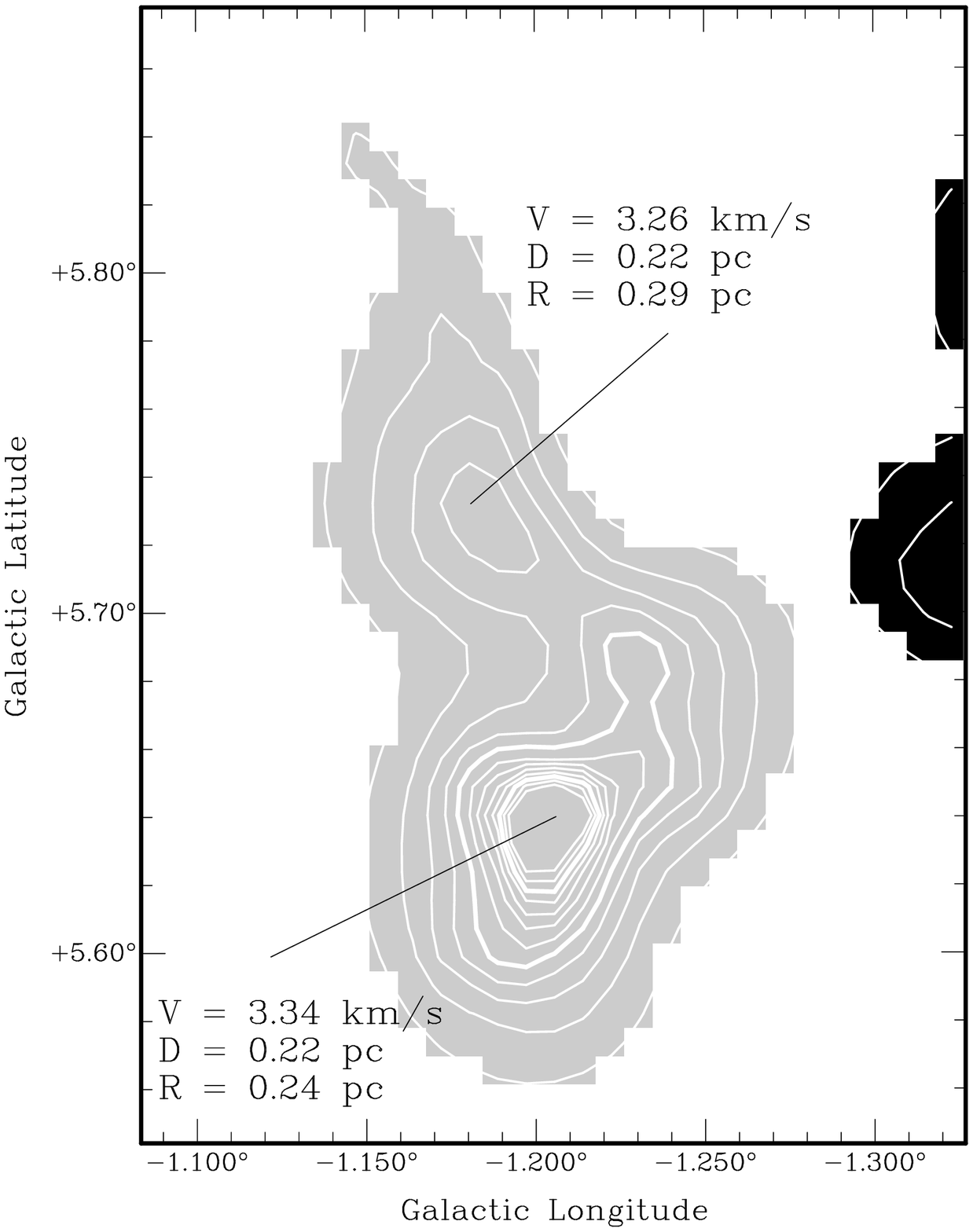}\\
\includegraphics[height=0.25\textheight,clip=true]{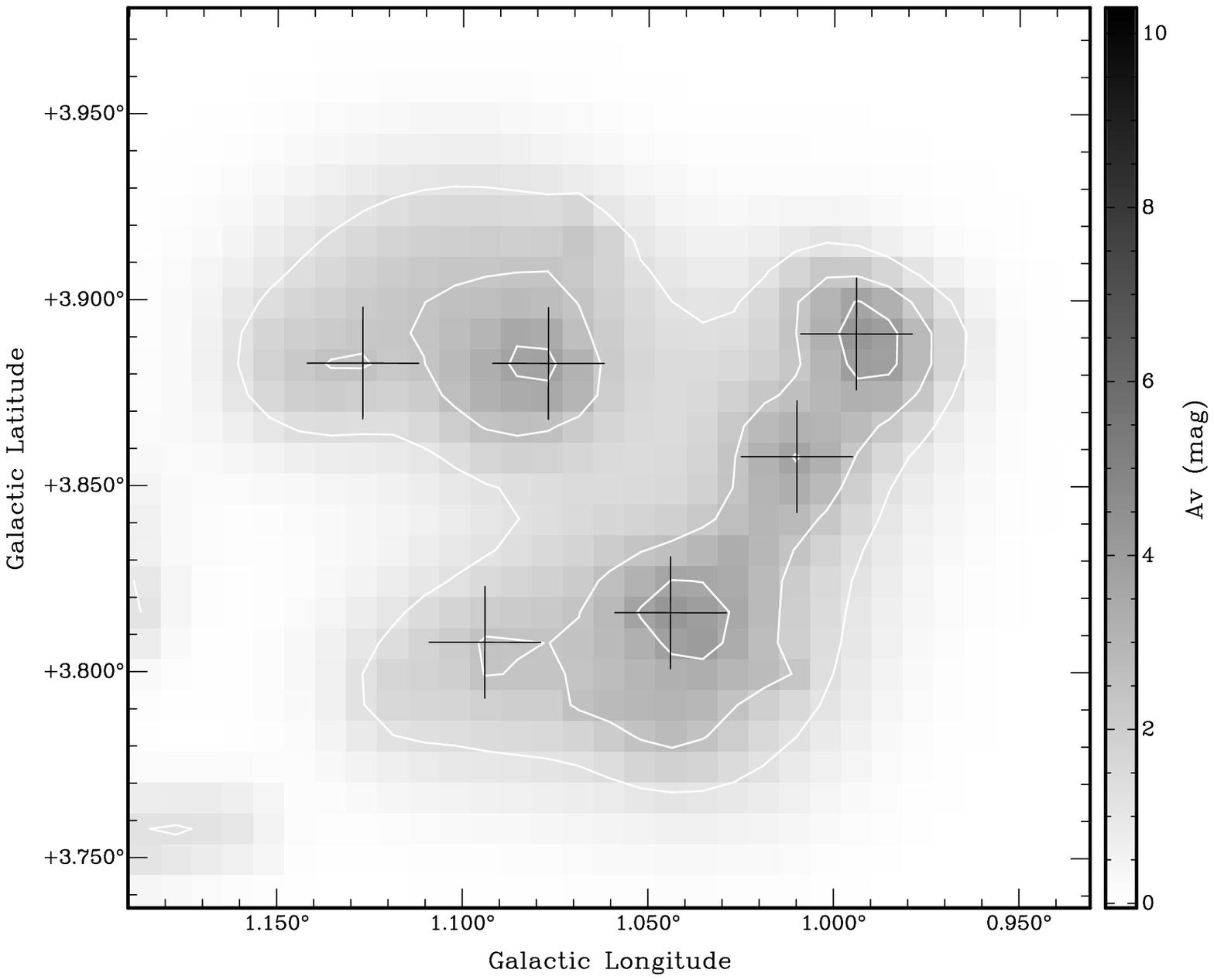} \includegraphics[height=0.25\textheight,clip=true]{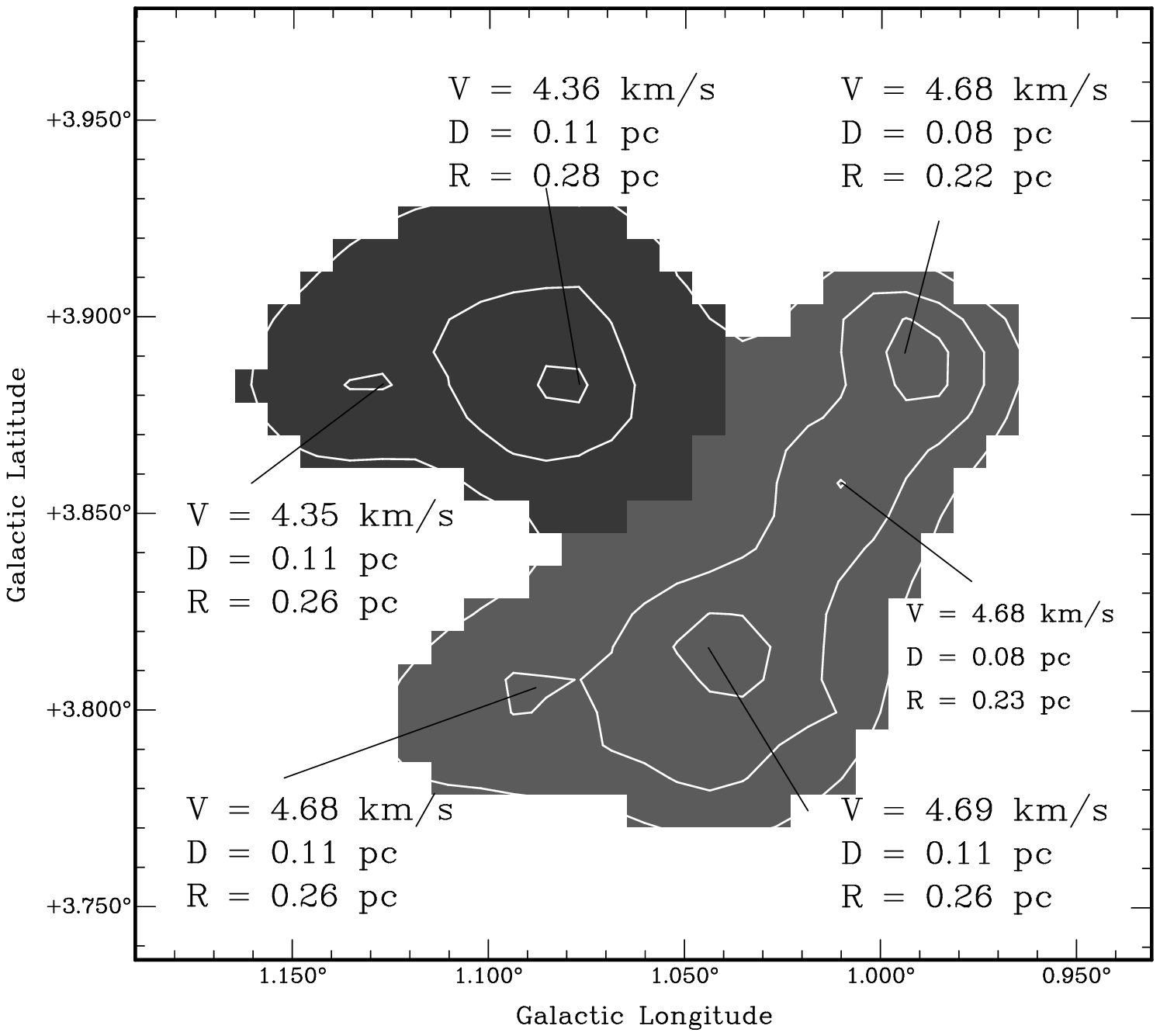}\\
\includegraphics[height=0.25\textheight,clip=true]{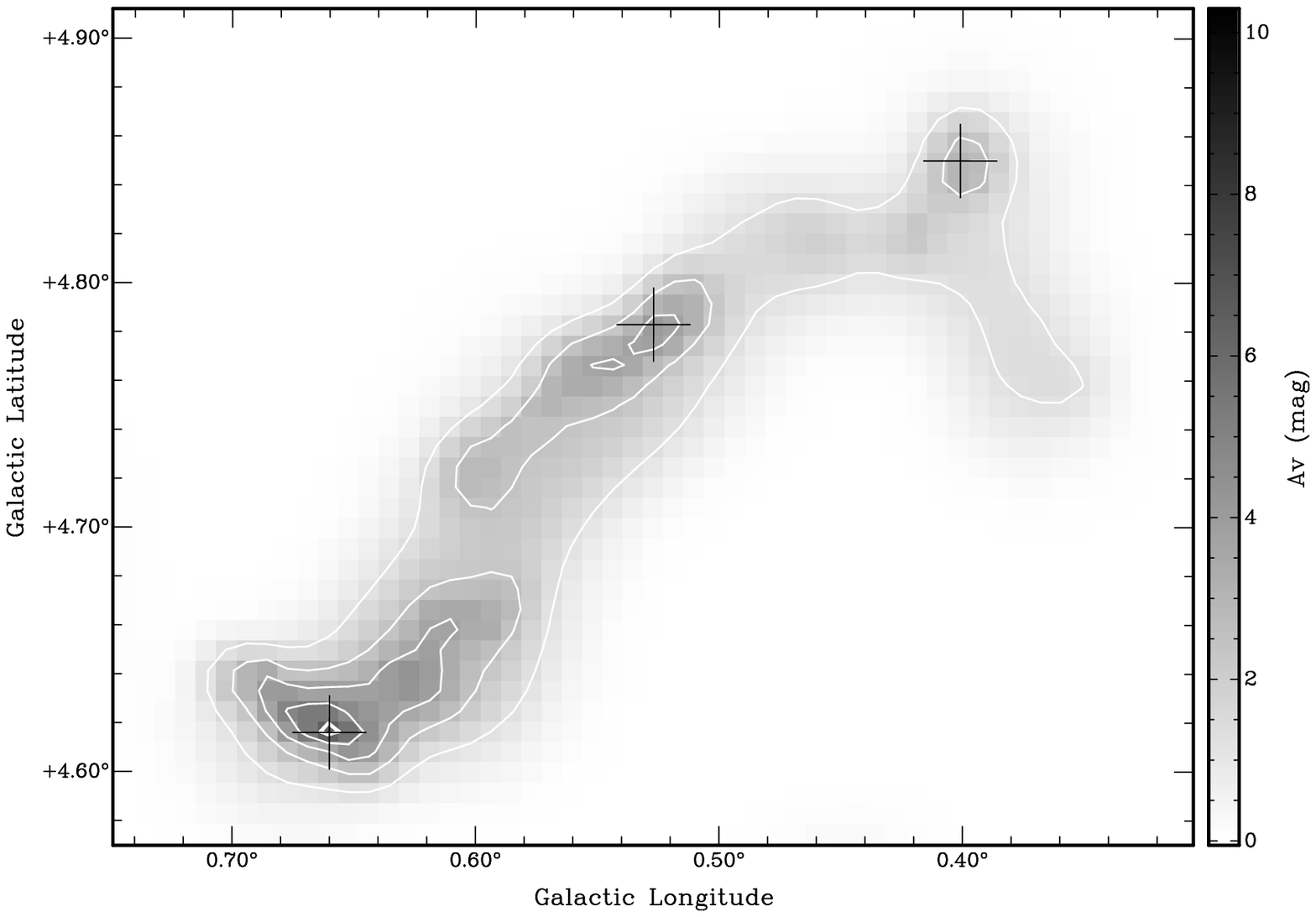} \includegraphics[height=0.25\textheight,clip=true]{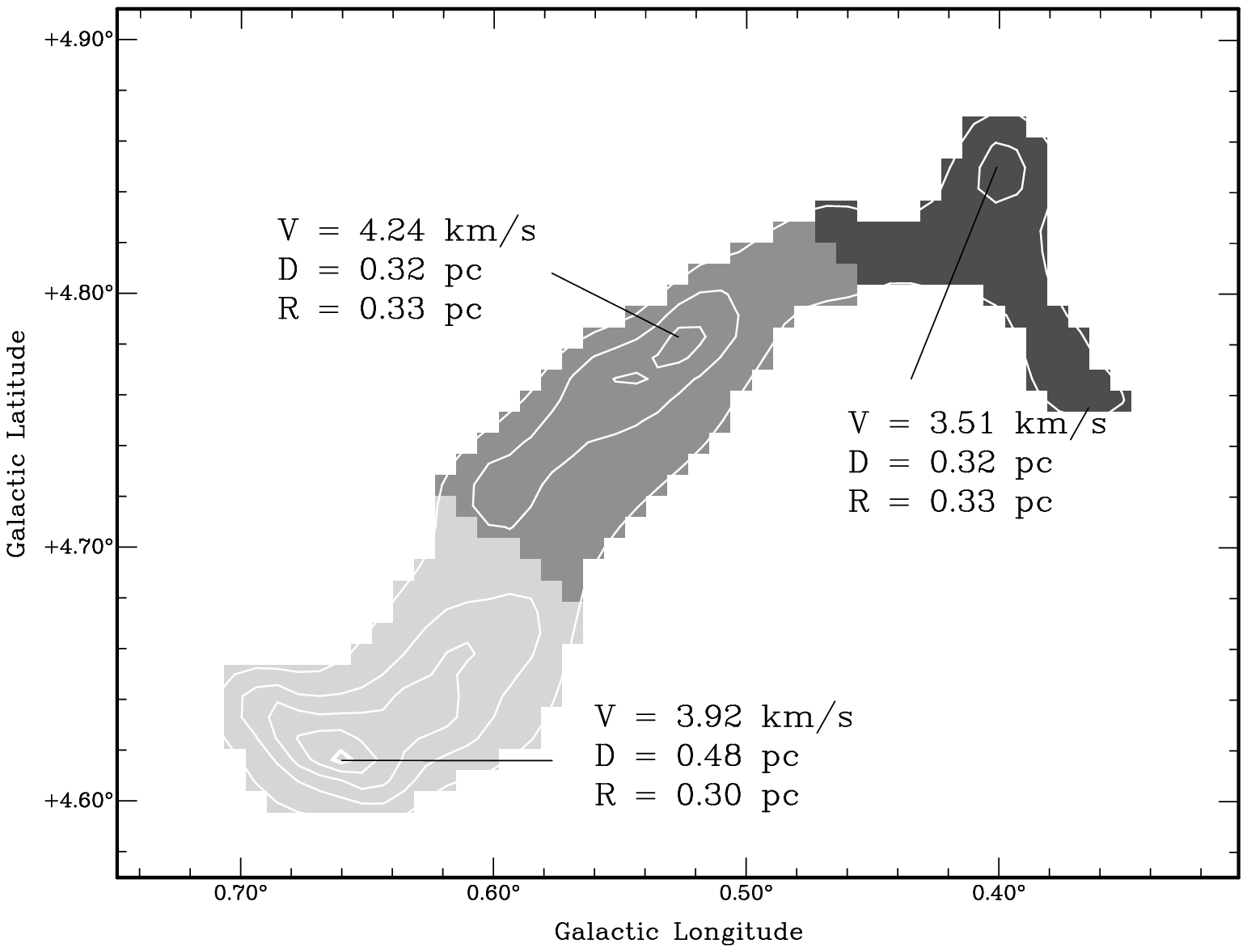}\\
\caption{\label{core-separation} Three examples of how we determine whether adjacent 
         extinction peaks are part of the same physical core or are
         unrelated.  The left panels show the background-subtracted
         extinction images with contours as defined in
         \S~\ref{clfind}. The crosses mark the positions of each
         extinction peak identified.  The right panels show the
         resulting cores after taking into account their velocity
         differences and Jeans lengths.  Marked on these images are
         the central velocities determined from the \ceo\, emission
         (V), the distance to the nearest extinction peak (D), and the
         Jeans length (R$_{J}$). In these images the color scale represents
         the area that is assigned to each core. In some cases we find
         that highly non-symmetric, complex extinction features have
         almost the same central velocity (top and middle panels).  In
         other cases, however, we find that adjacent extinction peaks
         within the same large scale extinction feature can have very
         different velocities (lower panel).}
\end{figure}

Figure~\ref{core-separation} shows three examples of how we determine
whether adjacent extinction peaks are physically associated. The left
panels show the background-subtracted extinction images with contours
as defined in \S~\ref{clfind}. The crosses mark the position of each
extinction peak identified.  The right panels show the resulting cores
after taking into account the velocity differences and Jeans lengths
of the extinction peaks.  Marked on these images are the central
velocities determined from the \ceo\, emission (V), the distance to
the nearest extinction peak (D), and the Jeans length (R$_{J}$). Using the
criteria, $\delta$V $<$ \cod\, and D $<$ R$_{J}$, we have determined if
adjacent extinction peaks are isolated entities or part of the same
physical core. In these images the color scale represents the area
that is assigned to each core.  In some cases we find that highly
non-symmetric, complex extinction features have almost the same
central velocity: these cores appear to have low-extinction tails
(e.g.\, top and middle panels in Fig.~\ref{core-separation}).  In
other cases, however, we find that adjacent extinction peaks within
the same large scale extinction feature can have very different
velocities (e.g.\, lower panels of Fig.~\ref{core-separation}). We
assume that these are physically differentiated or distinct cores.

\begin{figure}
\center
\includegraphics[angle=90,width=0.8\textwidth]{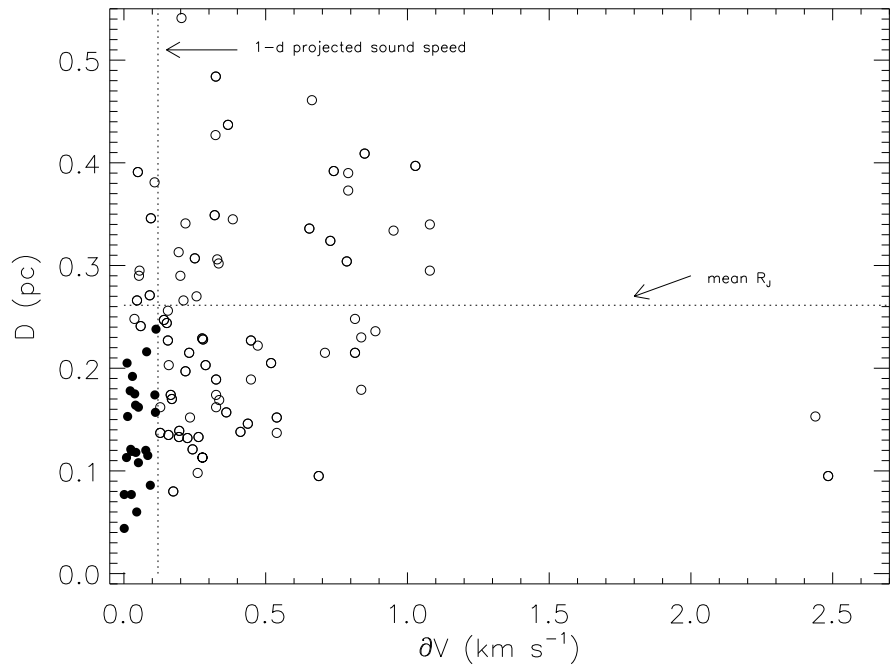}
\caption{\label{vel-diff} The velocity difference ($\delta$V) between each 
          extinction peak and all others in a larger extinction
          feature plotted against the distance to the extinction
          peak's nearest neighbor (D). The filled circles mark the
          $\delta$V and D for the extinction peaks that were merged
          into a single core. Forty one extinction peaks were found to be
          physically associated. These were merged into a total of 17
          cores for inclusion in the final list.}
\end{figure}

This procedure is summarized in Figure~\ref{vel-diff} which plots the
velocity difference ($\delta$V) between each extinction peak and every
other within each large extinction structure against the distance to
the extinction peak's nearest neighbor (D). The filled circles mark
the $\delta$V and D for extinction peaks that were determined to be
part of the same core. In total, we find that 41 extinction peaks have
$\delta$V $<$ \cod\, and D $<$ R$_{J}$. These were merged into 17
cores for inclusion in the final list. We note, however, that this
method will only allow us to separate cores that have significant
relative motion along the line-of-sight. If the relative motion
between adjacent cores is in fact in the plane of the sky, we will
have incorrectly merged them. Considering a core to core velocity
dispersion of $\sim$ 0.4\,\kms\, \citep{Gus-pipe}, we crudely estimate
that no more than 12 of the 41 cores may have been incorrectly merged.

To determine the masses of the merged cores, we calculate the
sum of the total extinction within the area associated with each
individual extinction peak. In all cases, we derive sizes using the
total number of pixels, converting this to a radius assuming that the
total area is a circle. For the remaining extinction peaks, we assume
that they are unrelated cores. Although many of these extinction peaks
have a neighbor that is either close in velocity or distance, they do
not satisfy both of these criteria. Thus, we assume these are simply
chance superpositions of unrelated cores along the line of sight.  To
determine their masses we use the total extinction directly output
from clumpfind.

\begin{figure}
\center
\includegraphics[angle=-90,width=0.9\textwidth]{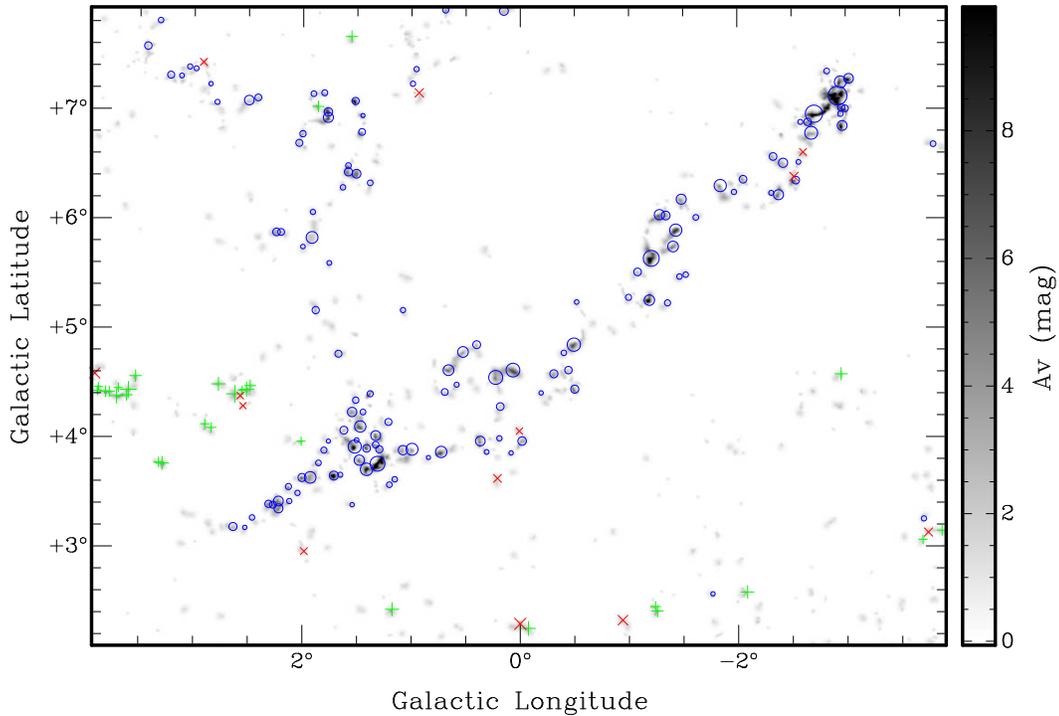}
\caption{\label{pipe-cores} The `cores-only', background-subtracted 2MASS visual extinction 
         (\av) image \citep{Lombardi06}. Using the \ceo\, emission
         toward each of the extinction peaks identified by clumpfind,
         we have determined which of the 201 extinction peaks are
         associated with dense gas, the Pipe Nebula, and if they are
         physically associated with any nearby extinction. Marked on
         this image are blue circles which correspond to the location
         and approximate extent of each of the 134 dense cores
         associated with the Pipe Nebula. The green plus symbols mark
         the extinction peaks that are associated with
         foreground/background molecular clouds.  The red crosses mark
         the extinction peaks that had no detectable \ceo\, emission.}
\end{figure}

Thus, after consideration of the velocity differences and Jeans
lengths of the identified extinction peaks, we find that there are 134
physically distinct dense cores associated with the Pipe
Nebula. Figure~\ref{pipe-cores} shows the location and approximate
extent of the cores identified within the Pipe Nebula.

\subsection{Core Mass Functions}

\begin{figure}
\center
\includegraphics[angle=90,width=0.9\textwidth]{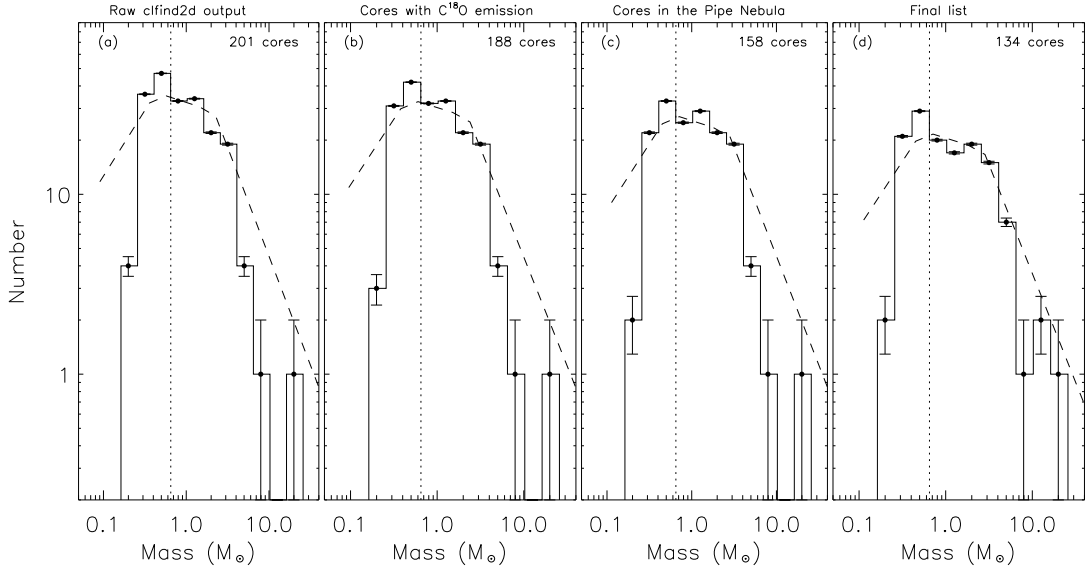}\\
\caption{\label{cmfs} Derived CMFs shown as binned histograms. The four panels
     correspond to different core samples when taking into
     consideration the \ceo\, molecular line emission. Note that the
     shape of the mass function changes considerably when we use the
     \ceo\, emission to guide the core extraction from the extinction
     image (i.e. between panels (a) and (d)).  The dashed line
     corresponds to the scaled field star IMF of \cite{Muench02}. For each panel
     we determine, via a $\chi ^{2}$ minimization, the offsets between the CMF and the IMF. 
     To accurately determine these scaling factors, we minimize the $\chi ^{2}$ between
     the distributions only for masses above the completeness limit.
     The derived parameters are summarized in 
     Table~\ref{CMF-IMF-comparison}. The vertical dotted line marks
     the mass completeness limit \citep{Jouni}. Included on the histograms 
     are the errors for each bin (calculated as the square
     root of the number per bin). }
\end{figure}

The derived CMFs are represented in Figure~\ref{cmfs} as binned histograms. 
Included on each plot is a scaled stellar IMF for the Trapezium cluster 
\citep{Muench02}. The vertical dotted line marks the mass completeness
limit calculated in the simulations of \cite{Jouni}.

The four panels in this figure represent the different stages in the
selection process for dense cores outlined above and illustrate the
changes in the shape of the CMF that occur as the result of the
refinement of the core sample as one includes more information from
the \ceo\, emission. Figure~\ref{cmfs}~(a) shows the derived CMF for
the 201 extinction peaks identified from the extinction image using
clfind2d and the contour levels described in \S~\ref{clfind}. This
represents the CMF derived using a blind application of clfind2d and
includes all extinction peaks as cores, regardless of whether or not
they have associated \ceo\, emission and, thus, dense gas. This will
include spurious and unrelated background cores and may also
artifically break high-mass cores into many lower-mass objects.

If we then consider only extinction peaks that have \ceo\, emission
and dense gas, the number of extinction peaks included in the CMF is
reduced to 188, the shape of which changes only slightly for the
lowest masses (M $<$ 2\,\Msun; Fig.~\ref{cmfs}~(b)). This is not
surprising considering the identification of spurious cores is a
significant effect only at the lowest extinction levels. Similarly,
our exclusion of cores lying outside the radial velocity range
considered for the Pipe Nebula (30 cores; Fig.~\ref{cmfs}~(c))
slightly alters the shape of the CMF for  masses $\lesssim$ 2\,\Msun.

A more significant change in the shape of the CMF occurs when we
consider the \ceo\, velocity differences and angular distances between
neighboring extinction peaks, as shown in Figure~\ref{cmfs}~(d). In
this case, extinction features are merged if their velocity
differences are less than the 1-dimensional projected sound speed and
separations are less than a Jeans length. This results in an increase
in the number of cores at the high-mass end (M $>$ 5\,\Msun) and a
decrease in the number of cores between masses of $\sim$0.3 and
$\sim$3\,\Msun.

Regardless of the exact shape, it appears that none of the CMFs shown
in Figure~\ref{cmfs} are characterized by a single power-law. Instead,
there is a break from a single power-law form providing a physical
scale or characteristic mass for the CMF around $\sim$
2--3\,\Msun. The lowest mass bins in the CMF are most likely seriously
effected by incompleteness. Thus, the position of the peak and the
turnover at the very lowest masses (M$\sim$ 0.4\,\Msun) may be
unreliable.  The simulations show that this mass range is most
affected by the incompleteness due to the wavelet decomposition. While
all the CMFs peak at roughly the same mass and show a break that is
well above the mass completeness limit, the most likely to be reliably
tracing the underlying distribution of core masses is the CMF shown in
Figure~\ref{cmfs}~(d). Thus, from here forward we adopt this distribution as the CMF for the Pipe Nebula.

It is of interest to core and star formation studies to compare the forms of the CMF and stellar IMF.  
To achieve this quantitatively, we have performed a $\chi ^{2}$ minimization between the 
Pipe CMF and the stellar IMF by simultaneously scaling the IMF in both the
x and y directions. The scaling factor in x direction will give the mass scaling between
the CMF and the IMF. Assuming that each core will give rise to a $\sim$1 star
on average, this offset will give an estimate of the star-formation
efficiency (SFE), that is, how much of the typical core mass is converted
into the final stellar mass. We have also calculated the $\chi ^{2}$ probability and
use this to estimate the errors. The quoted errors are
calculated from the range in the values for which the $\chi ^{2}$
probability is greater than 95\,\% (i.e. 2$\sigma$). We estimate 
that the SFE is 22 $\pm$ 8\,\% and that the break in the CMF occurs at a 
mass of 2.7 $\pm$ 1.3\,\Msun. Using the derived scaling factors for the IMF, we have also performed
a Kolmogorov-Smirnov (KS) test between the scaled-up stellar IMF and the Pipe CMF. 
We find that the probability that the distributions are derived from the same parent population
is 47\%.

For further comparison and to give a quantitative measure of how the inclusion of more 
information from the \ceo\, emission
effects the derived parameters, we have also performed the above analysis on 
the other three CMFs shown in Figure~\ref{cmfs}. Table~\ref{CMF-IMF-comparison} 
lists the derived parameters (mass scaling, SFE, and CMF break point) and their
errors for each panel of the CMFs shown in Figure~\ref{cmfs}.
Although Figures~\ref{cmfs} (a) and (b) contain spurious and unrelated cores, we find that the 
derived parameters are similar to those determined for the other distributions. 
Considering each of the Pipe cores that have \ceo\, emission as separate entities, as shown
 in Figure~\ref{cmfs} (c), we find that the CMF differs from the scaled 
IMF significantly for the highest masses. This is reflected in a low KS probability (7\%) that the two 
are derived from the same parent population. Indeed, all three of these distributions have 
significantly lower probabilities (7--8\%)
of being derived from the same parent distribution as the stellar IMF compared to the adopted Pipe CMF (47\%).

\section{Discussion}

\subsection{Derived core properties}

Because of the identification and selection methods used here, the
individual core properties may differ slightly from those listed in
our previous work \citep{Alves07,Gus-pipe,Lada-pipe,Rathborne-pipe}.
However, the mean values for the radii, n(\hh), and \sigmant\, are
identical to our previous work, i.e., mean radii of $\sim$0.09\,pc,
n(\hh) of $\sim$7.3$\times$10$^{3}$\,\cmc, and \sigmant\, of
$\sim$0.18\,\kms.  Although the properties of the individual cores may
differ, the ensemble properties remain unchanged: the cores appear to
be mostly pressure confined entities whose properties are determined
by the approximate balance between external and internal pressures
coupled with self-gravity.  Table~\ref{core-properties} lists the
derived properties for each core.

By comparing the input masses to those derived for the cores when
extracted using clfind2d, \cite{Jouni} determined the overall
uncertainty in the derived masses. There are three sources of error in
the mass calculation. They are due to (1) the image noise, (2) the
limiting extinction level considered by clumpfind, and (3) the degree
of core crowding. The uncertainty in the derived masses due to the
image noise can be determined via the dispersion in the ratio of the
derived mass to true mass. The simulations of \cite{Jouni} suggests
that the derived masses have an uncertainty of $\sim$ 25--30\%.
Because the clfind2d algorithm only includes pixels down to a limiting
extinction level (in this case an \av\,of 1.2\,mags), the core masses
will tend to be systematically underestimated. \cite{Jouni} show that
the extent to which the derived masses are underestimated varies with
core mass. For our clumpfind parameters, the simulations reveal that
there is no correction necessary for cores with masses $>$
2\,\Msun. For cores with masses $<$ 2\,\Msun, however, the derived
masses are typically 85\% of the true mass.

Masses may also be incorrectly determined if the cores are crowded and
overlapping. \cite{Jouni} defined a metric to describe core crowding
which considers both the cores' relative separation and their
diameter: f = mean core separation / mean core diameter. For cores
within the Pipe Nebula, we find that their mean separation is $\sim$
0.38 pc while their mean diameter is $\sim$ 0.18\,pc. The f-ratio is
$\sim$2.1, implying the crowding in the Pipe Nebula is minimal. Thus,
the error in the cores masses will not be dominated by the effect of
core crowding.

\subsection{The Core Mass Function for the Pipe Nebula}

\begin{figure}
\center
\includegraphics[angle=90,width=0.45\textwidth]{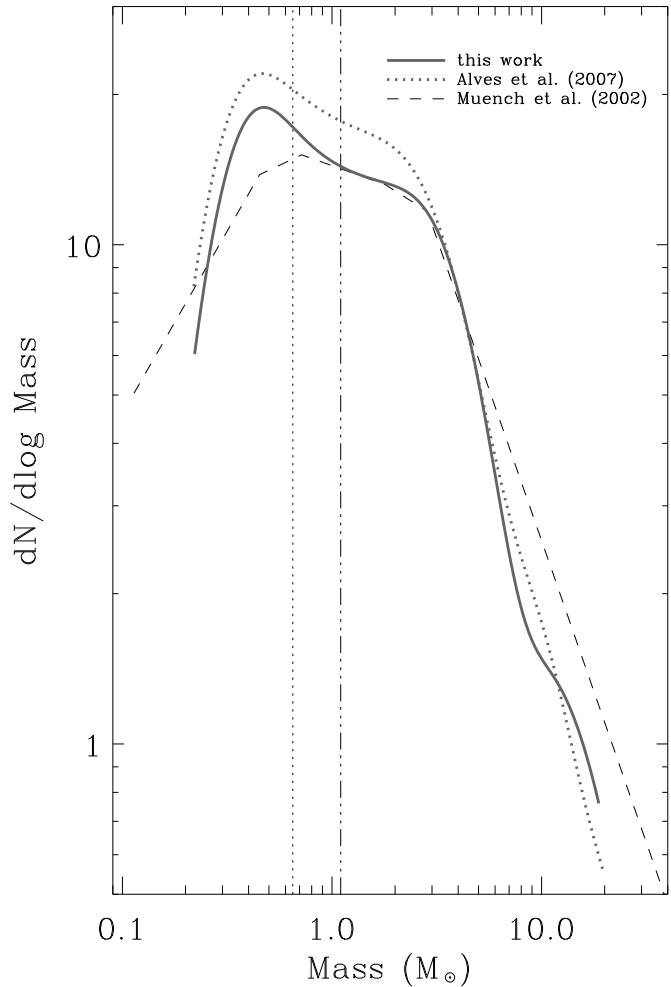}
\caption{\label{final-cmf}  The improved CMF for the Pipe Nebula as a 
         probability density function (solid curve) compared to the
         CMF of \cite{Alves07} (dotted curve).  The dashed line
         corresponds to the field star IMF of \cite{Muench02} scaled
         up by a factor of $\sim$ 4.5. We interpret this difference in
         scaling to be a measure of the star formation efficiency (22
         $\pm$ 8\%) that will likely characterize the dense core
         population in the cloud at the end of the star formation
         process.  The vertical dotted line marks the mass
         completeness limit, while the vertical dot-dashed line marks
         the fidelity limit \citep{Jouni}. We confirm earlier results
         that suggested that the Pipe CMF departs from a single
         power-law with a break at $\sim$ 2.7 $\pm$ 1.3\,\Msun.}
\end{figure}

The derived CMF for the Pipe Nebula is shown as a probability density
function in Figure~\ref{final-cmf} (solid curve). To generate the probability density
functions we use a Gaussian kernel of width 0.15 in units of log mass
\citep{Silverman86}. Because this does not require discrete data bins, it
more faithfully reproduces the detailed structure of the CMF compared
to a binned mass function.  Included on this
plot is the CMF from \cite{Alves07} (dotted curve) and the stellar IMF
for the Trapezium cluster \citep{Muench02} scaled in mass by a factor
of $\sim$ 4.5 (dashed line). The vertical dotted line marks the mass
completeness limit of $\sim$0.65\,\Msun. This limit gives the 90\%\,
mass completeness limit which reflects the core detections based on
the extinction sensitivity, wavelet decomposition, and input clumpfind
parameters. The vertical dotted-dashed line marks the fidelity limit
of $\sim$1.1\,\Msun, which corresponds to the mass above which the
precise shape of the CMF is reliable. This limit is calculated by
comparing the input and derived CMFs from the simulations and
basically accounts for the fact that some cores, while reliably
detected, have an incorrect mass due to the effects of core
overlaps. Both of these limits were calculated from the simulations of
\cite{Jouni}.

Overall, the shape of the CMF matches well with the scaled-up stellar
IMF. Similar to the result of \cite{Alves07}, we find that the CMF is
not characterized by a single power-law function. It can be described
as a power-law shape above a mass of $m_{break} \approx$ 2.7\,\Msun, at which point there is a break from the single
power-law and a clear flattening of the function towards lower
masses. The presence of such a break above the completeness limit is
significant because it imparts a scale or characteristic mass to the
CMF and enables a physically more meaningful comparison with the
stellar IMF.  The stellar IMF displays a similar break at
$m_{break}(\rm{IMF})\ \approx $ 0.6\, \Msun. The ratio of the two
characteristic masses gives a direct measure of the SFE.  As mentioned previously, we have
performed a $\chi ^{2}$ minimization between the stellar IMF and the Pipe CMF to give a
quantitative measure of the SFE. Assuming
that the dense cores in the Pipe will evolve to ultimately form stars,
we find a characteristic SFE of $\sim$ 22 $\pm$ 8\% for the
Pipe core population.

While the mass completeness limit is $\sim$ 0.65\,\Msun, all the cores
included in the CMF below this mass are real cores and are associated
with dense gas.  Although the second break and apparent
turnover of the CMF at M$\sim$ 0.4\,\Msun\, may be artificial due to
the incompleteness of our sample, these features are tantalizingly
close to a complementary second break and turnover in the scaled
stellar IMF. The possibility that these features may be similar in
both distributions potentially strengthens the connection between the
CMF and the stellar IMF.

Although the overall shapes of the Pipe CMF and stellar IMF are
similar, the CMF does appear to fall off more steeply toward higher
masses than the IMF.  If this is significant it would suggest that the
IMF is slightly wider than the CMF and would require the SFE to be an
increasing function of core mass in order to recover the stellar IMF
from the CMF in a simple one-to-one transformation.  However, given
the increased uncertainty due to the sampling errors and small number
statistics that characterize the higher mass bins, it is not possible
to argue with any degree of confidence that the two distributions are
different in this mass regime.


The general similarity in the shapes of the CMF and the IMF holds
significant implications for understanding one of the key unsolved
problems of star formation research, the origin of stellar mass. The
fundamental nature of stars and their evolution has long been well
explained by the theory of stellar structure and evolution developed
in the last century.  This theory has very successfully accounted for
the mass-dependent luminosities, sizes and lifetimes of stars on the
main sequence as well as their mass-dependent, post main-sequence
evolution. However this theory is silent on the question of how stars
get their masses in the first place. Because star formation research
has established that stars form in dense molecular cloud cores, the
similarity of the CMF and stellar IMF suggests that the IMF derives
its shape directly from the CMF. If in fact the shapes of the CMF and
IMF are the same, the derived SFE will apply to all core masses. This,
in turn, would suggest that the IMF results from a simple one-to-one
transformation of cores into stars. Thus, it may be possible to trace
the origin of the IMF directly to the CMF (modified by a star
formation efficiency) and the origin of stellar masses directly to the
origin dense core masses.

However, the concept of a simple one-to-one transformation of cores
into stars at a constant star formation efficiency is likely an
oversimplification of the actual star formation process which is
undoubtedly more complex. In particular, the more massive cores, that
is, those cores in the cloud whose masses exceed the critical
Bonnor-Ebert or Jeans mass are very likely to produce binary star
systems. Indeed, the most massive cores may even fragment and produce
small groups or clusters of stars. Thus, a strict one-to-one
correspondence between cores and individual stars can not be
preserved. However, the shape of the resulting IMF could be retained,
especially for higher stellar masses, if the SFE were to vary with
mass, being higher for higher mass cores, for example. Nonetheless,
recent simulations by \cite{Swift08} show that even when the internal
fragmentation of cores in a CMF is considered, the shape of the
resulting IMF is very similar to the shape of the input CMF (apart
from some details that may effect the very high and low mass ends in a
small, but measurable way). Moreover, their results indicate that even
if the SFE is not constant across the complete mass range, the
resulting IMFs are not that different in shape from the original CMF.
Thus, while a one-to-one correspondence between cores and stars may
not hold for all cores, the shape of the resulting IMF is likely to be
similar to the original shape of the CMF and a characteristic or mean
SFE can be measured by the ratio in the characteristic masses of the
two distributions.

\section{Summary and Conclusions}

In order to generate a more accurate CMF for the Pipe Nebula and thus
enable a more detailed and meaningful comparison between the CMF and
the stellar IMF, we have produced an improved census of the dense
cores and their properties within the Pipe Nebula.

Guided by the recent simulations of \cite{Jouni}, we re-examined the
extinction map for the Pipe Nebula and extracted an improved and more
reliable list of extinction peaks and candidate dense cores.  By
systematically observing each of these extinction peaks in \ceo\,
emission, we have identified which peaks are associated with dense gas
and which are associated with the Pipe molecular cloud. Moreover, we
employed information about the kinematic state of the gas derived from
our \ceo\ survey to refine the identifications of massive cores by
distinguishing internal core structure from apparent structure caused
by the overlap of physically discreet cores along similar
lines-of-sight.  The inclusion of spurious or unrelated cores, as well
as the incorrect merging or separation of extinction features, can
significantly affect the shape of the CMF constructed from
observations.  However, by combining measurements of visual extinction
and molecular-line emission these effects can be minimized and more
accurate measurements of core masses can be obtained.  With more
robust core identifications and masses we were able to construct an
improved CMF for the Pipe Nebula.

We confirm the earlier results which indicated a departure in the Pipe
CMF from a single power-law form.  We find the break point at a mass
of $\sim$ 2.7 $\pm$ 1.3\,\Msun, well above the mass
completeness and fidelity limits for the observed core sample.
Moreover, similar to Alves et al. (2007) we find that the overall
shape of the CMF is generally similar to the stellar IMF except for a
difference in mass scaling of about a factor of $\sim$ 4.5. We
interpret this difference in scaling to be a measure of the star
formation efficiency (22 $\pm$ 8\%) that will likely
characterize the dense core population in the cloud at the end of the
star formation process. These dense cores comprise an invaluable
catalog for follow-up studies of the connection between core and star
formation.

\acknowledgments

We thank the referee for a thorough reading of the paper and the
useful suggestions which have improved the paper. These
observations were supported through NASA Origins grant NAG-13041 and
the NASA Spitzer GO program (PID 20119) and supported by JPL contract
1279166.



\begin{deluxetable}{lcccccccccc}
\small
\tablecolumns{10}
\tablewidth{0pt}
\tablecaption{\label{core-properties} Core properties}
\tablehead{\colhead{Core} & \multicolumn{2}{c}{Coordinates} & \colhead{Peak \av} & \colhead{R} & \colhead{M} & \colhead{n(\hh)} & \multicolumn{3}{c}{\ceo\, emission} & \colhead{A07\tablenotemark{a}} \\
\colhead{} & \colhead{$\ell$} & \colhead{$b$} & \colhead{} & \colhead{} & \colhead{} & \colhead{} & \colhead{\tastar} & \colhead{\vlsr} & \colhead{$\Delta$V} & \colhead{core} \\
\colhead{} & \colhead{(deg)} & \colhead{(deg)} & \colhead{(mag)} & \colhead{(pc)} & \colhead{(\Msun)} & \colhead{(10$^{4}$\, \cmc)} & \colhead{(K)} & \colhead{(\kms)} & \colhead{(\kms) }& }
\startdata
       1  &    -3.80  &     6.68  &      2.0  &     0.06  &    0.5  &      0.9  &     0.13  &     5.49  &     0.39  &   2   \\
       2\tablenotemark{b}  &    -3.70  &     3.25  &      2.0  &     0.05  &    0.4  &      1.0  &     0.16  &     3.94  &     0.66  &   4   \\
       3\tablenotemark{b}  &    -3.03  &     7.28  &     10.9  &     0.10  &    3.0  &      1.3  &     1.78  &     3.61  &     0.28  &   6   \\
       4\tablenotemark{b}  &    -2.99  &     7.01  &      8.8  &     0.07  &    1.5  &      1.9  &     0.88  &     3.97  &     0.52  &   7   \\
       5\tablenotemark{b}  &    -2.97  &     6.85  &     10.0  &     0.10  &    3.1  &      1.2  &     1.62  &     3.57  &     0.35  &   8   \\
       6  &    -2.96  &     7.01  &      7.3  &     0.07  &    1.9  &      2.0  &     1.29  &     3.80  &     0.61  &   7   \\
       7  &    -2.95  &     7.25  &      8.1  &     0.12  &    4.0  &      0.9  &     1.52  &     3.45  &     0.44  &   12 11   \\
       8\tablenotemark{b}  &    -2.95  &     6.96  &      3.8  &     0.06  &    0.6  &      1.4  &     0.28  &     3.54  &     1.05  &   7   \\
       9  &    -2.93  &     7.12  &     20.2  &     0.19  &   19.4  &      1.1  &     1.17  &     3.64  &     0.86  &   12 7 11   \\
      10\tablenotemark{b}  &    -2.83  &     7.35  &      2.8  &     0.06  &    0.6  &      1.1  &     0.50  &     3.75  &     0.32  &   13   \\
      11  &    -2.71  &     6.96  &     15.6  &     0.18  &   12.7  &      0.9  &     2.21  &     3.62  &     0.35  &   12 14 15   \\
      12  &    -2.68  &     6.78  &      4.0  &     0.13  &    3.3  &      0.6  &     1.25  &     3.35  &     0.24  &   16   \\
      13\tablenotemark{b}  &    -2.65  &     6.88  &      4.9  &     0.08  &    1.2  &      1.2  &     2.06  &     3.39  &     0.39  &   15   \\
      14  &    -2.58  &     6.88  &      2.2  &     0.05  &    0.3  &      1.0  &     1.15  &     3.49  &     0.37  &   18   \\
      15  &    -2.56  &     6.52  &      2.1  &     0.05  &    0.3  &      1.1  &     0.24  &     3.07  &     0.65  &   19   \\
      16  &    -2.54  &     6.35  &      7.1  &     0.08  &    1.7  &      1.4  &     1.11  &     3.70  &     0.26  &   20   \\
      17  &    -2.42  &     6.51  &      3.4  &     0.09  &    1.6  &      0.8  &     0.76  &     3.55  &     0.35  &   21   \\
      18  &    -2.38  &     6.22  &      5.7  &     0.10  &    2.4  &      0.9  &     1.04  &     3.64  &     0.33  &   23 22   \\
      19  &    -2.33  &     6.57  &      3.3  &     0.08  &    1.1  &      0.9  &     0.65  &     3.40  &     0.42  &   21   \\
      20\tablenotemark{b}  &    -2.31  &     6.24  &      3.2  &     0.05  &    0.5  &      1.4  &     0.59  &     3.28  &     0.33  &   23   \\
      21  &    -2.05  &     6.36  &      3.8  &     0.08  &    1.1  &      0.9  &     0.47  &     3.74  &     0.36  &   25   \\
      22  &    -1.97  &     6.25  &      2.0  &     0.05  &    0.4  &      1.0  &     0.39  &     2.92  &     0.34  &   26   \\
      23  &    -1.84  &     6.30  &      4.6  &     0.13  &    3.1  &      0.7  &     0.95  &     3.18  &     0.19  &   27   \\
      24  &    -1.77  &     2.57  &      3.0  &     0.04  &    0.3  &      1.6  &     0.16  &     5.27  &     1.80  &   28   \\
      25  &    -1.62  &     6.01  &      1.6  &     0.06  &    0.4  &      0.8  &     0.34  &     3.48  &     0.24  &   29   \\
      26  &    -1.52  &     5.49  &      2.0  &     0.06  &    0.4  &      0.9  &     0.35  &     3.23  &     0.35  &   30   \\
      27  &    -1.48  &     6.18  &      4.1  &     0.10  &    2.0  &      0.7  &     0.58  &     3.38  &     0.60  &   31   \\
      28\tablenotemark{b}  &    -1.46  &     5.47  &      2.3  &     0.06  &    0.4  &      1.0  &     0.75  &     3.14  &     0.32  &   32   \\
      29  &    -1.43  &     5.90  &      8.1  &     0.13  &    4.3  &      0.9  &     1.96  &     3.36  &     0.39  &   33   \\
      30  &    -1.41  &     5.75  &      5.3  &     0.11  &    2.7  &      0.8  &     2.11  &     3.11  &     0.28  &   34   \\
      31  &    -1.36  &     5.23  &      2.0  &     0.06  &    0.5  &      0.9  &     0.17  &     2.93  &     0.49  &   35   \\
      32\tablenotemark{b}  &    -1.34  &     6.03  &      4.2  &     0.09  &    1.7  &      0.9  &     0.67  &     3.50  &     0.47  &   36   \\
      33  &    -1.28  &     6.04  &      6.6  &     0.11  &    3.1  &      1.1  &     1.49  &     3.31  &     0.36  &   37 39   \\
      34\tablenotemark{b}  &    -1.21  &     5.64  &     19.7  &     0.17  &    9.3  &      0.8  &     2.01  &     3.34  &     0.29  &   40   \\
      35\tablenotemark{b}  &    -1.19  &     5.26  &     17.9  &     0.11  &    3.9  &      1.3  &     2.22  &     3.91  &     0.25  &   42 41   \\
      36  &    -1.08  &     5.52  &      2.6  &     0.08  &    0.9  &      0.7  &     1.10  &     3.37  &     0.33  &   43   \\
      37  &    -1.00  &     5.28  &      2.1  &     0.06  &    0.5  &      0.9  &     0.44  &     3.42  &     0.37  &   44   \\
      38  &    -0.52  &     5.24  &      1.9  &     0.05  &    0.3  &      1.1  &     0.41  &     3.09  &     0.32  &   46   \\
      39  &    -0.50  &     4.44  &      5.8  &     0.08  &    1.4  &      1.0  &     1.18  &     2.91  &     0.34  &   47   \\
      40  &    -0.49  &     4.85  &      7.0  &     0.14  &    4.2  &      0.7  &     2.98  &     3.61  &     0.31  &   48   \\
      41  &    -0.44  &     4.62  &      2.5  &     0.08  &    0.9  &      0.8  &     0.95  &     3.60  &     0.43  &   49   \\
      42  &    -0.40  &     4.77  &      1.9  &     0.06  &    0.4  &      1.0  &     0.67  &     3.90  &     0.24  &   50   \\
      43  &    -0.31  &     4.58  &      3.9  &     0.08  &    1.2  &      0.9  &     1.67  &     3.65  &     0.36  &   51   \\
      44  &    -0.19  &     4.41  &      1.9  &     0.04  &    0.2  &      1.2  &     0.87  &     3.56  &     0.36  &   52   \\
      45\tablenotemark{b}  &    -0.02  &     3.97  &      3.5  &     0.09  &    1.4  &      0.9  &     0.80  &     5.84  &     0.29  &   54   \\
      46  &     0.07  &     4.62  &      7.4  &     0.14  &    5.6  &      0.8  &     1.55  &     3.61  &     0.49  &   56   \\
      47  &     0.08  &     3.86  &      2.3  &     0.05  &    0.3  &      1.3  &     0.28  &     5.86  &     0.90  &   57   \\
      48  &     0.15  &     7.91  &      2.8  &     0.09  &    1.3  &      0.7  &     0.26  &     3.72  &     0.17  &  -   \\
      49  &     0.18  &     4.28  &      2.0  &     0.08  &    0.8  &      0.7  &     0.29  &     3.77  &     0.71  &   58   \\
      50\tablenotemark{b}  &     0.19  &     3.99  &      2.2  &     0.05  &    0.4  &      1.0  &     0.67  &     4.74  &     0.29  &   59   \\
      51  &     0.23  &     4.55  &      4.9  &     0.15  &    4.6  &      0.6  &     1.04  &     3.66  &     0.34  &   62 61   \\
      52  &     0.31  &     3.87  &      2.7  &     0.05  &    0.4  &      1.3  &     0.50  &     5.44  &     0.37  &   63   \\
      53  &     0.37  &     3.97  &      5.5  &     0.10  &    2.1  &      0.9  &     1.58  &     4.99  &     0.56  &   65 64 66   \\
      54  &     0.40  &     4.85  &      3.0  &     0.08  &    0.9  &      0.7  &     0.69  &     3.51  &     0.34  &   67   \\
      55  &     0.53  &     4.78  &      4.0  &     0.11  &    2.2  &      0.7  &     1.46  &     4.24  &     0.45  &   67   \\
      56  &     0.59  &     4.48  &      2.0  &     0.05  &    0.4  &      1.1  &     0.59  &     4.46  &     0.40  &   68   \\
      57  &     0.66  &     4.62  &      6.4  &     0.11  &    2.8  &      0.8  &     1.91  &     3.91  &     0.46  &   70 69   \\
      58  &     0.69  &     7.91  &      4.0  &     0.06  &    0.7  &      1.2  &     0.83  &     6.09  &     0.29  &   72   \\
      59  &     0.69  &     4.42  &      2.3  &     0.07  &    0.7  &      0.8  &     0.52  &     4.26  &     0.32  &   73   \\
      60  &     0.73  &     3.87  &      6.9  &     0.12  &    3.0  &      0.8  &     1.96  &     4.21  &     0.32  &   74   \\
      61\tablenotemark{b}  &     0.84  &     3.82  &      2.4  &     0.04  &    0.3  &      1.3  &     0.65  &     5.14  &     0.47  &   75   \\
      62  &     0.96  &     7.37  &      2.3  &     0.05  &    0.4  &      1.0  &     0.22  &     6.27  &     0.71  &   77   \\
      63  &     0.99  &     7.24  &      1.9  &     0.05  &    0.3  &      0.9  &     0.22  &     6.27  &     0.80  &   78   \\
      64  &     0.99  &     3.89  &      4.2  &     0.12  &    3.1  &      0.7  &     0.75  &     4.68  &     0.73  &   79 80   \\
      65  &     1.08  &     3.88  &      3.8  &     0.10  &    1.7  &      0.7  &     0.93  &     4.36  &     0.41  &   79 80   \\
      66  &     1.08  &     5.17  &      2.4  &     0.05  &    0.4  &      1.1  &     0.37  &     3.45  &     0.32  &   81   \\
      67  &     1.15  &     3.62  &      2.2  &     0.06  &    0.4  &      1.0  &     0.31  &     6.36  &     0.49  &   82   \\
      68  &     1.20  &     3.57  &      2.6  &     0.06  &    0.5  &      1.1  &     1.09  &     6.30  &     0.27  &   84   \\
      69  &     1.21  &     4.14  &      2.3  &     0.07  &    0.7  &      0.8  &     0.30  &     4.68  &     0.69  &   85   \\
      70  &     1.29  &     3.89  &      3.9  &     0.07  &    1.1  &      1.2  &     0.55  &     5.23  &     0.32  &   88   \\
      71  &     1.31  &     3.76  &     15.0  &     0.15  &   10.6  &      1.2  &     0.77  &     4.44  &     0.58  &   87   \\
      72  &     1.33  &     3.93  &      4.3  &     0.07  &    0.9  &      1.3  &     0.98  &     5.47  &     0.48  &   88   \\
      73  &     1.33  &     4.02  &      6.3  &     0.10  &    2.5  &      1.0  &     1.50  &     4.45  &     0.45  &   89 86   \\
      74  &     1.38  &     4.40  &      6.3  &     0.06  &    1.1  &      1.8  &     0.74  &     4.28  &     0.34  &   91   \\
      75  &     1.38  &     6.33  &      2.2  &     0.06  &    0.5  &      1.0  &     0.27  &     4.98  &     0.51  &   90   \\
      76  &     1.41  &     3.71  &     10.2  &     0.13  &    4.6  &      0.9  &     1.23  &     5.20  &     0.35  &   93   \\
      77  &     1.41  &     3.90  &      7.0  &     0.08  &    1.6  &      1.4  &     1.62  &     5.18  &     0.39  &   92   \\
      78  &     1.45  &     4.23  &      2.7  &     0.06  &    0.5  &      1.1  &     0.88  &     4.10  &     0.25  &   97   \\
      79  &     1.45  &     6.95  &      4.3  &     0.04  &    0.4  &      2.2  &     0.54  &     4.76  &     0.49  &   95   \\
      80  &     1.46  &     6.80  &      5.2  &     0.07  &    1.1  &      1.3  &     0.29  &     4.77  &     0.66  &   96   \\
      81  &     1.47  &     4.10  &      7.0  &     0.12  &    3.6  &      0.9  &     0.60  &     4.35  &     0.70  &   97   \\
      82  &     1.48  &     3.79  &      6.3  &     0.11  &    2.4  &      0.8  &     1.17  &     5.35  &     0.50  &   98 94   \\
      83  &     1.50  &     3.97  &      2.6  &     0.05  &    0.4  &      1.3  &     0.44  &     4.64  &     0.75  &   102   \\
      84  &     1.51  &     6.41  &      5.6  &     0.09  &    2.3  &      1.3  &     1.47  &     4.71  &     0.28  &   99   \\
      85  &     1.51  &     4.34  &      2.2  &     0.07  &    0.6  &      0.9  &     0.15  &     3.81  &     0.81  &   100   \\
      86  &     1.52  &     7.08  &     12.0  &     0.07  &    1.9  &      1.9  &     1.65  &     3.34  &     0.26  &   101   \\
      87  &     1.52  &     3.92  &     11.4  &     0.14  &    5.4  &      0.9  &     2.08  &     4.84  &     0.41  &   102   \\
      88  &     1.54  &     3.38  &      2.1  &     0.04  &    0.3  &      1.3  &     1.25  &     2.83  &     0.30  &   103   \\
      89  &     1.55  &     4.23  &      3.9  &     0.10  &    1.8  &      0.8  &     0.17  &     4.25  &     1.49  &   97   \\
      90  &     1.58  &     6.43  &      5.3  &     0.08  &    1.6  &      1.2  &     1.27  &     4.55  &     0.33  &   105   \\
      91  &     1.58  &     6.49  &      5.8  &     0.06  &    0.8  &      1.8  &     0.98  &     4.77  &     0.30  &   106   \\
      92  &     1.62  &     4.07  &      2.6  &     0.08  &    1.0  &      0.8  &     0.51  &     3.99  &     0.65  &   102   \\
      93  &     1.63  &     6.29  &      2.7  &     0.06  &    0.5  &      1.1  &     0.58  &     4.91  &     0.15  &   107   \\
      94  &     1.65  &     3.66  &      3.6  &     0.05  &    0.6  &      1.7  &     1.61  &     6.03  &     0.28  &   109   \\
      95  &     1.67  &     4.76  &      2.8  &     0.07  &    0.8  &      0.9  &     1.24  &     3.30  &     0.37  &   108   \\
      96  &     1.71  &     3.65  &     12.2  &     0.09  &    3.1  &      1.6  &     1.50  &     5.87  &     0.31  &   109   \\
      97\tablenotemark{b}  &     1.76  &     5.60  &      2.3  &     0.05  &    0.4  &      1.2  &     0.51  &     6.09  &     0.58  &   110   \\
      98  &     1.76  &     3.96  &      1.7  &     0.04  &    0.2  &      1.1  &     0.30  &     3.60  &     0.68  &   111   \\
      99  &     1.77  &     6.93  &      7.6  &     0.10  &    2.8  &      1.1  &     1.35  &     4.95  &     0.31  &   112 116   \\
     100  &     1.77  &     6.98  &      8.0  &     0.09  &    2.4  &      1.5  &     1.16  &     4.68  &     0.35  &   113   \\
     101\tablenotemark{b}  &     1.80  &     3.88  &      3.0  &     0.06  &    0.6  &      1.0  &     0.62  &     3.99  &     0.43  &   115   \\
     102  &     1.80  &     7.15  &      3.3  &     0.06  &    0.6  &      1.2  &     0.28  &     4.98  &     0.28  &   114   \\
     103  &     1.85  &     3.76  &      3.9  &     0.06  &    0.6  &      1.2  &     0.54  &     3.74  &     0.97  &   118   \\
     104  &     1.88  &     5.16  &      3.5  &     0.07  &    0.9  &      0.9  &     0.47  &     5.24  &     0.50  &   119   \\
     105  &     1.90  &     7.15  &      2.7  &     0.06  &    0.5  &      1.0  &     0.29  &     4.70  &     0.23  &   114   \\
     106  &     1.91  &     6.06  &      2.6  &     0.05  &    0.4  &      1.1  &     0.36  &     4.49  &     0.85  &   120   \\
     107  &     1.92  &     5.83  &      2.4  &     0.12  &    2.2  &      0.5  &     0.31  &     4.96  &     0.32  &   121   \\
     108  &     1.93  &     3.63  &      4.2  &     0.12  &    2.9  &      0.7  &     1.28  &     3.77  &     0.30  &   123 122   \\
     109  &     2.00  &     5.75  &      1.6  &     0.05  &    0.3  &      1.0  &     0.29  &     4.99  &     0.34  &   125   \\
     110  &     2.00  &     3.63  &      6.2  &     0.08  &    1.5  &      1.1  &     0.95  &     3.60  &     0.41  &   127   \\
     111  &     2.00  &     6.78  &      3.6  &     0.06  &    0.7  &      1.1  &     0.46  &     4.26  &     0.13  &   126   \\
     112  &     2.04  &     6.70  &      3.1  &     0.07  &    0.8  &      0.9  &     0.30  &     4.77  &     0.64  &   126   \\
     113  &     2.04  &     3.49  &      2.0  &     0.05  &    0.4  &      1.0  &     0.26  &     3.63  &     0.40  &   129   \\
     114  &     2.12  &     3.41  &      3.4  &     0.05  &    0.5  &      1.4  &     1.29  &     3.60  &     0.30  &   132   \\
     115  &     2.13  &     3.55  &      4.7  &     0.06  &    0.8  &      1.2  &     0.45  &     3.87  &     0.51  &   130   \\
     116  &     2.20  &     5.88  &      2.4  &     0.07  &    0.8  &      0.9  &     0.45  &     5.64  &     0.20  &   133   \\
     117  &     2.22  &     3.35  &      5.1  &     0.09  &    2.0  &      1.1  &     1.17  &     3.51  &     0.45  &   131   \\
     118  &     2.22  &     3.41  &      5.5  &     0.10  &    2.5  &      1.0  &     1.05  &     4.05  &     0.35  &   132   \\
     119  &     2.24  &     5.88  &      3.8  &     0.08  &    1.2  &      1.0  &     1.88  &     3.16  &     0.40  &   133   \\
     120  &     2.27  &     3.38  &      5.1  &     0.07  &    1.4  &      1.5  &     0.87  &     3.92  &     0.42  &   132 131   \\
     121  &     2.31  &     3.39  &      5.1  &     0.08  &    1.2  &      1.2  &     1.40  &     3.23  &     0.41  &   132 131   \\
     122  &     2.42  &     7.11  &      2.7  &     0.07  &    0.8  &      0.9  &     0.36  &     3.11  &     0.33  &   134   \\
     123  &     2.46  &     3.26  &      3.1  &     0.06  &    0.4  &      1.0  &     1.25  &     2.93  &     0.40  &   135   \\
     124  &     2.50  &     7.08  &      2.5  &     0.10  &    1.5  &      0.6  &     0.19  &     2.90  &     0.54  &   134   \\
     125  &     2.53  &     3.17  &      2.5  &     0.05  &    0.3  &      1.3  &     0.89  &     2.48  &     0.29  &   137   \\
     126  &     2.63  &     3.18  &      3.0  &     0.08  &    1.1  &      0.8  &     0.51  &     3.07  &     0.39  &   140   \\
     127  &     2.79  &     7.07  &      2.4  &     0.05  &    0.4  &      1.2  &     0.20  &     2.92  &     0.32  &   142   \\
     128  &     2.85  &     7.23  &      2.9  &     0.05  &    0.3  &      1.4  &     0.31  &     5.44  &     0.26  &   144   \\
     129  &     2.99  &     7.37  &      2.1  &     0.05  &    0.4  &      1.1  &     0.19  &     2.56  &     0.34  &   147   \\
     130  &     3.04  &     7.39  &      2.5  &     0.05  &    0.5  &      1.1  &     0.55  &     2.48  &     0.23  &   148   \\
     131  &     3.12  &     7.30  &      3.2  &     0.05  &    0.4  &      1.2  &     0.97  &     2.95  &     0.43  &   149   \\
     132  &     3.22  &     7.31  &      2.7  &     0.07  &    0.8  &      0.9  &     0.55  &     3.05  &     0.43  &   150   \\
     133  &     3.31  &     7.81  &      2.5  &     0.06  &    0.5  &      1.1  &     0.45  &     2.47  &     0.26  &   152   \\
     134  &     3.43  &     7.58  &      2.5  &     0.08  &    0.8  &      0.8  &     0.31  &     2.41  &     0.29  &   153   \\

\enddata
\tablenotetext{a}{Core number from the catalog of \cite{Alves07}. Thirty three 
     cores from the \cite{Alves07} catalog are not included here
     because they either had no associated \ceo\, emission or the
     \ceo\, emission was not in the range 1.3\,$<$ \vlsr $< $
     6.4\,\kms. Because of the differences in the way the cores were
     extracted from the extinction image and the fact that we merged
     or separated the extinction based on the associated \ceo\,
     emission, many of the cores from \cite{Alves07} are listed
     multiple times in our catalog. To be associated with a core in
     the current catalog, we require the cores of \cite{Alves07} to
     have more than 20 pixels contained within the boundary of the new
     core. This criteria was necessary to prevent entries where the
     cores within the two catalogs overlapped only at the very edges.}
\tablenotetext{b}{Extinction peaks were merged within this core.}
\end{deluxetable}
\begin{deluxetable}{lcccccc}
\small
\tablecolumns{10}
\tablewidth{0pt}
\tablecaption{\label{CMF-IMF-comparison} Derived quantites from the $\chi ^{2}$ minimization between the CMFs and IMFs in Figure~\ref{cmfs}.}
\tablehead{\colhead{Panel} & \colhead{Mass scaling} & \colhead{SFE} &\colhead{CMF break point} & \colhead{KS probability} \\
\colhead{} & \colhead{} & \colhead{(\%)} & \colhead{(\Msun)} & \colhead{(\%)} } 
\startdata
(a)                        &  3.6 $\pm$ 1.6 & 28 $\pm$ 15       & 2.2 $\pm$ 1.6             & 8  \\
(b)                        &  3.8 $\pm$ 1.6 & 26 $\pm$ 13       & 2.3 $\pm$ 1.6             & 7  \\
(c)                        &  4.5 $\pm$ 0.8 & 23 $\pm$ 4        & 2.7 $\pm$ 0.8             & 7  \\
(d)                        &  4.5 $\pm$ 1.3 & 22 $\pm$ 8        & 2.7 $\pm$ 1.3             & 47 \\
\enddata
\end{deluxetable}



\end{document}